\documentclass[prb,twocolumn,10pt,aps,showpacs,amsmath,amsfonts,floatfix,letterpaper,superscriptaddress]{revtex4-1}

\usepackage{times}
\usepackage{txfonts}

\usepackage{amsmath}
\usepackage{amssymb}
\usepackage{hyperref}
\usepackage{multirow}

\newcommand{\ee}{\mathrm{e}}
\newcommand{\ii}{\mathrm{i}}

\newcommand{\del}{\boldsymbol{\nabla}}
\newcommand{\Av}{\boldsymbol{A}}

\newcommand{\deltav}{\boldsymbol{\delta}}
\newcommand{\zerov}{\boldsymbol{0}}

\newcommand{\parder}[2]{\frac{\partial #1}{\partial #2}}
\newcommand{\parderat}[3]{\left(\frac{\partial #1}{\partial #2}\right)_{\!#3}}

\newcommand{\vphiv}{\boldsymbol{\varphi}}

\newcommand{\primecubic}[1]{#1\sub{\smash{cubic}}^{\smash{{\prime}}}}
\newcommand{\noprimecubic}[1]{#1\sub{\smash{cubic}}^{\smash{{\phantom{\prime}}}}}

\newcommand{\Bv}{\boldsymbol{B}}
\newcommand{\rv}{\boldsymbol{r}}
\newcommand{\mv}{\boldsymbol{m}}

\newcommand{\scE}{\mathcal{E}}
\newcommand{\scL}{\mathcal{L}}
\newcommand{\goz}{\mathfrak{z}}
\newcommand{\sigmam}{\boldsymbol{\sigma}}
\newcommand{\scZ}{\mathcal{Z}}
\newcommand{\scB}{\mathcal{B}}
\newcommand{\scG}{\mathcal{G}}

\newcommand{\beq}[1]{\begin{equation}\label{#1}}
\newcommand{\eeq}{\end{equation}}
\newcommand{\refeq}[1]{Eq.~(\ref{#1})}

\newcommand{\refeqand}[2]{Eqs.~(\ref{#1}) and (\ref{#2})}
\newcommand{\refcite}[1]{Ref.~\onlinecite{#1}}
\newcommand{\refcites}[1]{Refs.~\onlinecite{#1}}
\newcommand{\reffig}[1]{Fig.~\ref{#1}}
\newcommand{\reffigand}[2]{Figs.~\ref{#1} and \ref{#2}}
\newcommand{\refsec}[1]{Section~\ref{#1}}
\newcommand{\refsecand}[2]{Sections~\ref{#1} and \ref{#2}}
\newcommand{\refsecs}[2]{Sections~\ref{#1}--\ref{#2}}
\newcommand{\reftab}[1]{Table~\ref{#1}}

\newcommand{\punc}[1]{\,{\text{#1}}}
\newcommand{\sub}[1]{_{\text{#1}}}

\newcommand{\blp}{\boldsymbol{(}}
\newcommand{\brp}{\boldsymbol{)}}

\DeclareMathOperator{\Div}{div}

\usepackage{color}

\usepackage{graphicx}
\usepackage{ifpdf}
\ifpdf

\newcommand{\putinscaledfigure}[1]{\begin{center}\includegraphics[width=\columnwidth]{#1.pdf}\end{center}}
\else

\newcommand{\putinscaledfigure}[1]{\begin{center}\includegraphics[width=\columnwidth]{#1.eps}\end{center}}
\fi

\begin{document}

\title{Critical behavior in the cubic dimer model at nonzero monomer density}
\author{G. J. Sreejith}
\affiliation{Nordita, KTH Royal Institute of Technology and Stockholm University, Roslagstullsbacken 23, SE-106 91 Stockholm, Sweden}
\author{Stephen Powell}
\affiliation{Nordita, KTH Royal Institute of Technology and Stockholm University, Roslagstullsbacken 23, SE-106 91 Stockholm, Sweden}
\affiliation{School of Physics and Astronomy, The University of Nottingham, Nottingham, NG7 2RD, United Kingdom}

\begin{abstract}

We study critical behavior in the classical cubic dimer model (CDM) in the presence of a finite density of monomers. With attractive interactions between parallel dimers, the monomer-free CDM exhibits an unconventional transition from a Coulomb phase to a dimer crystal. Monomers acts as charges (or monopoles) in the Coulomb phase and, at nonzero density, lead to a standard Landau-type transition. We use large-scale Monte Carlo simulations to study the system in the neighborhood of the critical point, and find results in agreement with detailed predictions of scaling theory. Going beyond previous studies of the transition in the absence of monomers, we explicitly confirm the distinction between conventional and unconventional criticality, and quantitatively demonstrate the crossover between the two. Our results also provide additional evidence for the theoretical claim that the transition in the CDM belongs in the same universality class as the deconfined quantum critical point in the $\mathrm{SU}(2)$ JQ model.

\end{abstract}


\maketitle

\section{Introduction}
\label{SecIntroduction}

According to the Landau paradigm, the critical behavior occurring near a second-order phase transition can be understood through the long-wavelength fluctuations of the order parameter.\cite{Landau} Cases where such a picture does not correctly describe the critical properties, termed non-Landau or simply ``unconventional'' transitions, present an interesting extension of the established theory of thermal and quantum criticality.

Recent interest in this idea was sparked by the theoretical prediction of such a quantum phase transition in certain two-dimensional antiferromagnets.\cite{Senthil} In these systems, a continuous transition of the Landau paradigm is excluded by the presence of distinct order parameters in the two adjacent phases. It has been argued that the critical behavior is instead governed by ``spinons'', quasiparticles carrying spin of $\frac{1}{2}\hbar$, that are deconfined only at the critical point.\cite{Senthil} Evidence for the existence of such ``deconfined quantum criticality'' has been provided by quantum Monte Carlo studies of certain spin models.\cite{SandvikJQ,Lou,Sandvik,Pujari,Harada}

A conventional Landau transition is also excluded for some thermal transitions, when the ``disordered'' high-temperature phase has some kind of topological order.\cite{Alet} This can occur in constrained statistical systems, and leads to a second set of unconventional transitions that are, on the surface, only remotely connected to deconfined quantum criticality. Such examples are particularly interesting when the low-temperature phase has broken symmetry, and so the transition separates phases with distinct types of order, topological above the critical temperature and symmetry-breaking below.

It is remarkable that unconventional criticality of the latter variety is exhibited by one of the simplest conceivable constrained classical systems, the close-packed dimer model on the cubic lattice.\cite{Alet} This model describes the statistics of ``dimers'', objects that occupy pairs of neighboring sites of a lattice. In a close-packed dimer model, the allowed configurations are constrained so that every site is covered by exactly one dimer; we also consider here the case when ``monomer'' defects, where a site's occupation number deviates from unity, are permitted with a large but finite energy cost $\Delta$.

In the constrained limit, $\Delta = \infty$, and with no other interactions, the cubic dimer model (CDM) exhibits a liquid-like phase with strong correlations but no static order.\cite{Huse,HenleyReview} This can be interpreted as the Coulomb phase of an effective lattice gauge theory, in which the monomers act as charges (``monopoles''). A pair of test monomers inserted into the constrained system is subject to an effective interaction, induced by the fluctuations of the dimers. In the Coulomb phase, this interaction obeys a Coulomb law for large separation, implying that the defects can be separated to infinity with finite energy cost, and are hence deconfined.

In the presence of additional interactions between the dimers, a transition can occur into a phase where monomers are instead confined, by an effective potential proportional to their separation. With attractive interactions $v_2$ between neighboring parallel dimers, the phase at low temperature $T$ has columnar order [see \reffig{FigPhases}(a)] that breaks the rotation and translation symmetries of the lattice.\cite{Alet,Misguich} While a conventional order parameter can be identified, the critical behavior at this ``confinement transition'' is modified by the fluctuations of the gauge theory, in ways that cannot be described by a conventional Landau theory.\cite{HenleyReview,CubicDimers,Charrier,Chen}

In this work, we study the critical properties of this transition beyond the constrained limit. Reducing $\Delta/T$ from $\infty$ allows a finite density of monomers in the dimer model, screening the long-range interactions between test monomers and rendering the confinement criterion invalid. The confinement transition at $\Delta = \infty$ is therefore replaced by a conventional phase transition, described by Landau theory. For sufficiently large $\Delta/T$, however, the behavior is still strongly influenced by the presence of the nearby confinement critical point. We study this behavior using large-scale Monte Carlo simulations, to test predictions of scaling theory.

In particular, previous work\cite{MonopoleScalingPRL,MonopoleScalingPRB} has presented general theoretical arguments that the fugacity $z = \ee^{-\Delta/T}$ is a relevant scaling field under the renormalization group (RG) at the fixed point corresponding to the confinement transition. This result provides quantitative predictions for the behavior at nonzero monomer density, including for example, the shape of the phase boundary at $z > 0$. These applications of scaling theory have previously been tested in certain confinement transitions in spin ice,\cite{MonopoleScalingPRL,MonopoleScalingPRB} in which spontaneous symmetry breaking was absent. The present case is more demanding computationally, but also displays considerably richer phenomena, with symmetry breaking coinciding with deconfinement at the transition, and the phase boundary extending to $z>0$.

This alternative perspective is complementary to those exploited in previous Monte Carlo studies of the CDM, which have been restricted to the fully constrained limit.\cite{Alet,Misguich,Charrier,Chen,Papanikolaou,Charrier2} Our numerical results are entirely consistent with both the qualitative and quantitative predictions of scaling theory. They demonstrate a clear distinction between the critical behavior at $z=0$ and $z>0$, and thereby provide a particularly stringent test of the interpretation of the transition through confinement.

An additional interesting feature of the CDM transition is that its universality class is expected to be the same as that of the original deconfined critical point in quantum spin models.\cite{Senthil} This allows us to compare our results for this transition with those for the critical point in the JQ model.\cite{SandvikJQ} (Limited comparisons have already been made using results at $z=0$.\cite{Charrier2}) Agreement is again found, within the limits of accuracy that are inherent in such comparisons. Since the evidence for a continuous, non-Landau transition in the CDM is substantial, these results provide quite strong support for the theoretical scenario of deconfined quantum criticality.

\subsection*{Outline}

In \refsec{SecModelPhaseDiagram}, we review the cubic dimer model and its phase structure, first in terms of the original dimer degrees of freedom and then through the language of an effective gauge theory and the associated concept of confinement. We then focus, in \refsec{SecPhaseTransitions}, on the phase transitions from the dimer crystal, reviewing the critical theories for the transitions and then, in \refsecs{SecScalingConfinement}{SecFluxStiffness}, presenting a number of predictions for behavior in their neighborhood. These are tested in detail in \refsec{SecResults}, using numerical results from large-scale Monte Carlo simulations of the dimer model. We conclude in \refsec{SecConclusions} with a restatement of our main results and a brief discussion of their significance for the dimer model and for unconventional transitions more broadly. Some details of the Monte Carlo algorithm are given in an Appendix.

\section{Model and phase diagram}
\label{SecModelPhaseDiagram}

\subsection{Cubic dimer model}
\label{SecModel}

We consider a classical statistical model in which the elementary degrees of freedom are dimers, defined on the links of a cubic lattice. Labeling each site by a vector $\rv$ of integers, one can define an occupation number $d_\mu(\rv) \in \{0,1\}$, giving the number of dimers on the link joining $\rv$ and $\rv + \deltav_\mu$, where $\mu \in \{x,y,z\}$ and $\deltav_\mu$ is a unit vector. The number of dimers touching site $\rv$ is given by
\beq{EqDefinen}
n(\rv) = \sum_\mu [d_\mu(\rv) + d_\mu(\rv - \deltav_\mu)]\punc{.}
\eeq
In a close-packed hard-core dimer model, one constrains $n(\rv) = 1$ on every site $\rv$; instead, we also allow monomers, where a site has either no dimer or multiple dimers. We are interested in the case where the elementary defects, with $n(\rv) = 0$ or $2$, cost a large energy $\Delta$, and so occur at low density. Larger deviations, where $n(\rv) > 2$, are expected to be negligible near the transition, and are excluded. Defining the on-site energy $\scE(n)$, we therefore set $\scE(1) = 0$, $\scE(0)=\scE(2)=\Delta$, and $\scE(n) = \infty$ otherwise.

The transition to an ordered state is driven by an attractive interaction between pairs of nearest-neighbor parallel dimers,\cite{Alet} whose number is
\beq{EqDefineN2}
N_2 = \sum_{\rv} \sum_{\mu,\nu \neq \mu} d_\mu(\rv) d_\mu(\rv + \deltav_{\nu}) \punc{.}
\eeq
It is also necessary (see \refcites{Papanikolaou,Charrier2} and the final paragraph of \refsec{SecPhaseStructure}) to include additional interactions, and we follow Charrier and Alet \cite{Charrier2} by using a term that counts the number of parallel dimers around cubes of the lattice,
\beq{EqDefineN4}
N_4 = \sum_{\rv} \sum_{\substack{\mu,\nu \neq \mu\\\rho \neq \mu,\nu}} d_\mu(\rv) d_\mu(\rv + \deltav_{\nu})  d_\mu(\rv + \deltav_\rho) d_\mu(\rv + \deltav_{\nu} + \deltav_\rho)
\punc{.}
\eeq

The full configuration energy can then be written as
\beq{EqConfigEnergy}
\begin{aligned}
&E = \sum_{\rv} \scE\blp n(\rv)\brp + E\sub{dimer}\\
&E\sub{dimer} = v_2 N_2 + v_4 N_4\punc{.}
\end{aligned}
\eeq
We are interested in the case of attractive interactions, $v_2 < 0$, and in the following we choose units where $v_2 = -1$.

The partition function is given by
\beq{EqPartitionFunction}
\scZ = \sum_{\{d_\mu(\rv)\}} \ee^{-E/T}\punc{.}
\eeq
We use a lattice with periodic boundary conditions and an even number $L$ of sites in each direction.

\subsection{Phase structure}
\label{SecPhaseStructure}

\begin{figure*}
\begin{center}
\begin{tabular}{ccc}
\includegraphics[width=0.25\textwidth]{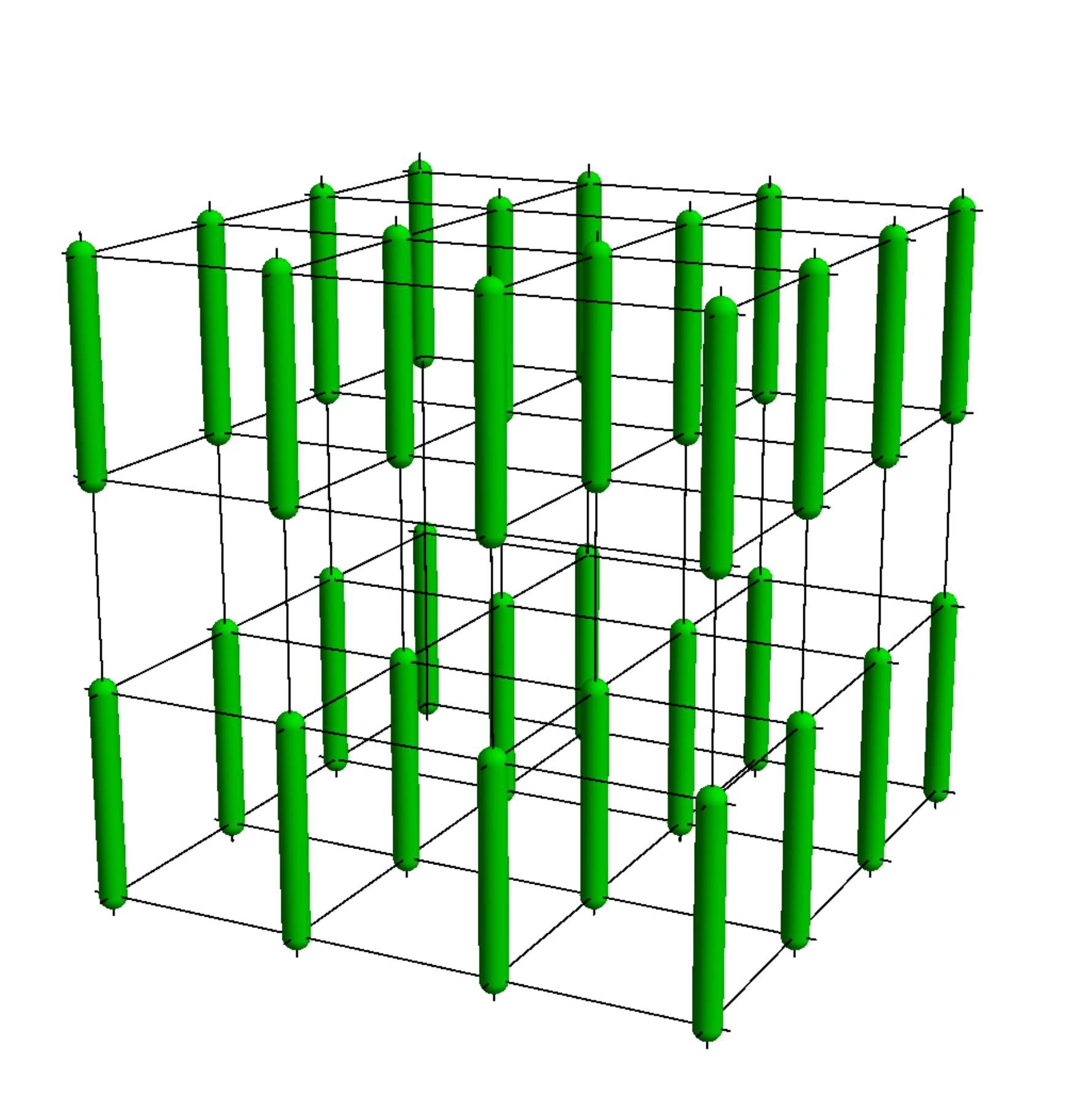}&
\includegraphics[width=0.25\textwidth]{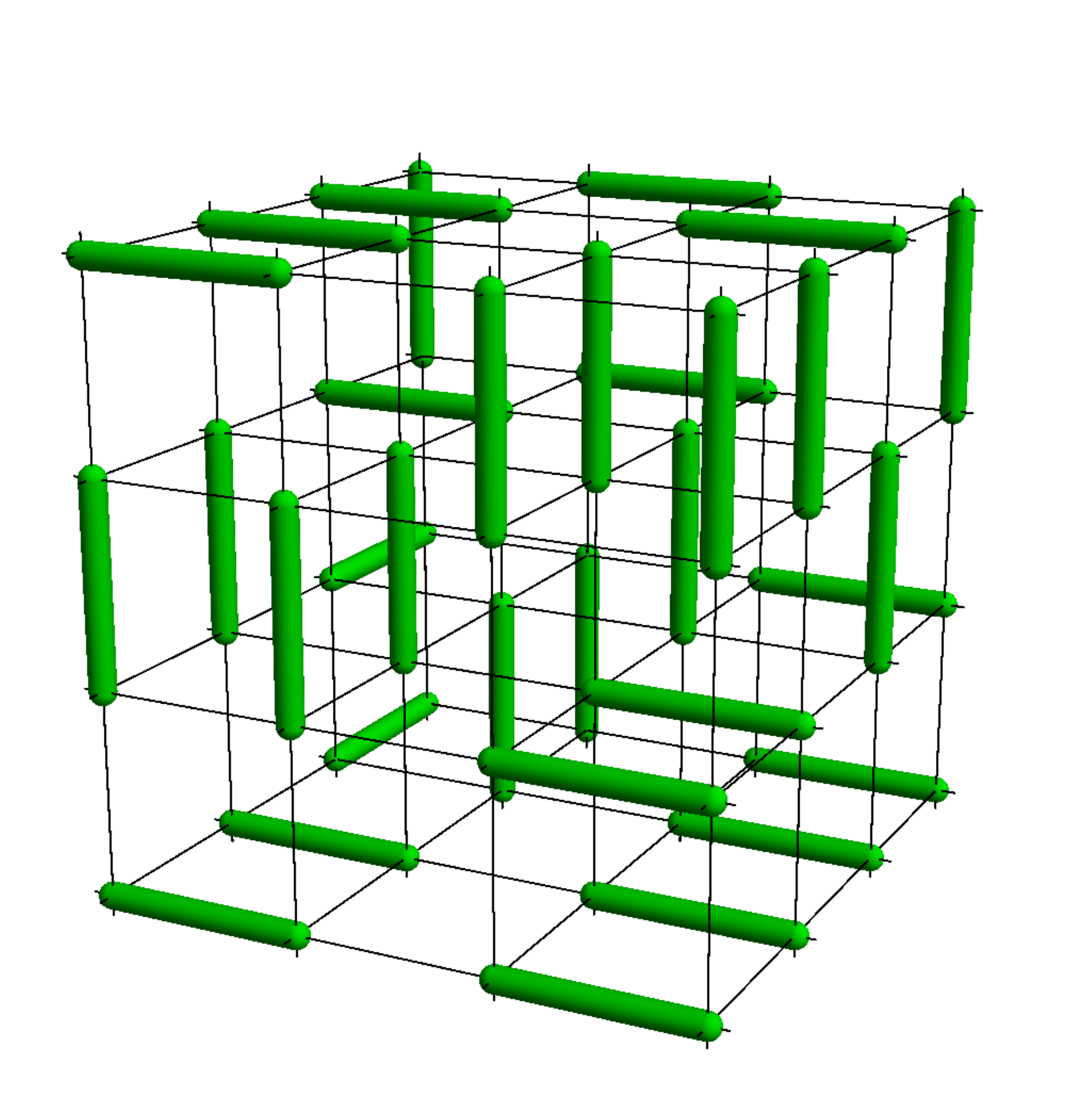}&
\includegraphics[width=0.25\textwidth]{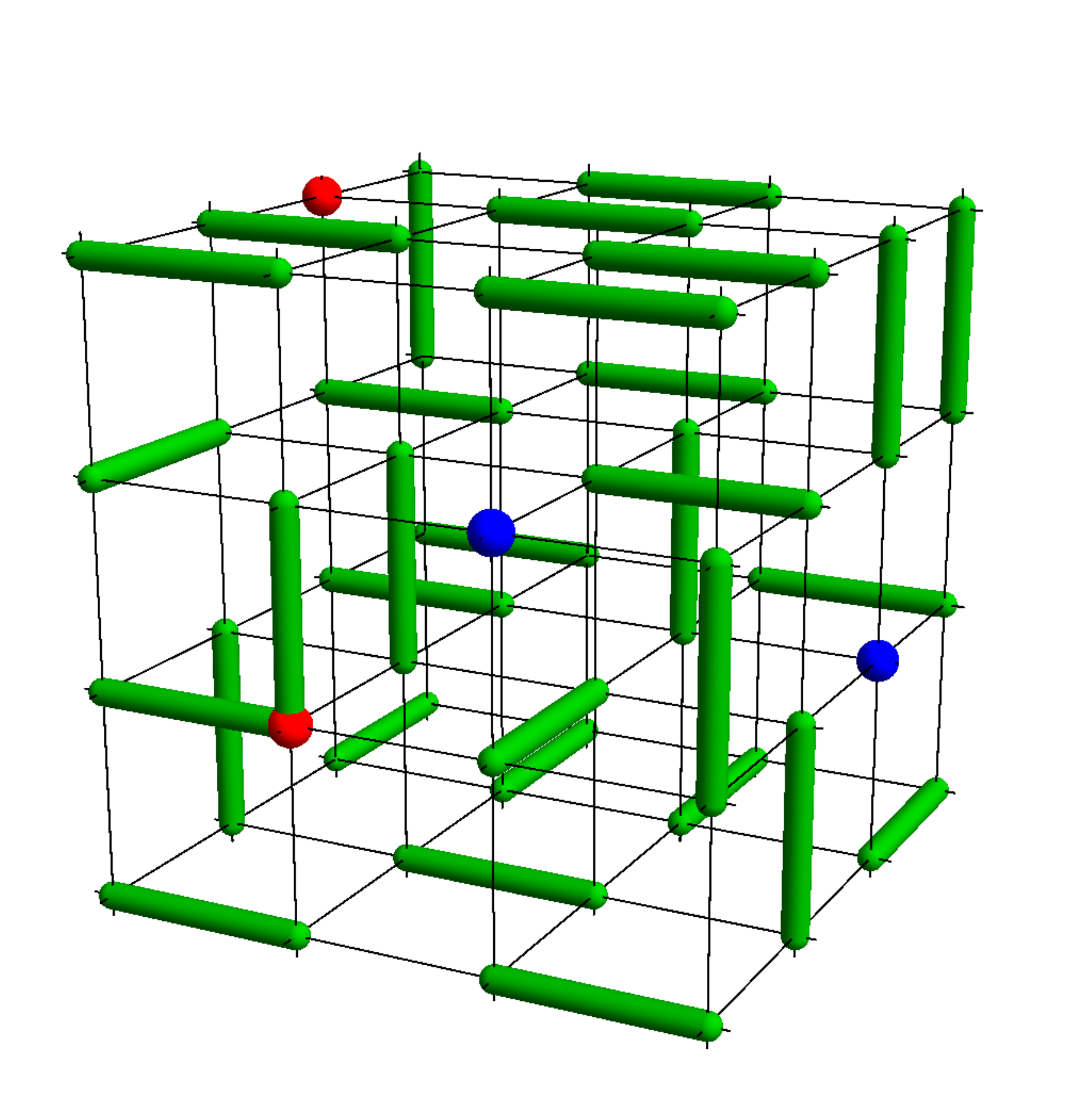}\\
(a)&(b)&(c)
\end{tabular}
\end{center}
\caption{Illustrations of three phases of dimers (green) occupying the links of a cubic lattice. (a) Columnar dimer crystal, with all dimers aligned with the vertical direction and arranged in layers, maximizing the number $N_2$ of plaquettes with parallel dimers. (b) Coulomb phase, where dimers are not ordered, but are constrained so that the dimer number $n(\rv)$ is fixed to unity for every site of the lattice. (c) Thermally disordered phase, with a nonzero density of defects in the dimer constraint. Sites with zero or two dimers cost energy $\Delta$, and act as charges in the effective gauge theory. The sign of the charge (indicated with red and blue spheres) depends on both the sign of the deviation, $n(\rv) - 1$, and on the sublattice.
\label{FigPhases}}
\end{figure*}

Considering first the case where $v_4 = 0$, the energy is globally minimized by a columnar dimer crystal, as shown in \reffig{FigPhases}(a).
The dimers are aligned with a particular cubic axis, and form sheets in which each dimer has four parallel neighbors. There are six such configurations---one example has $d_\mu(\rv)=1$ for $\mu=z$ and $r_z$ even, and $d_\mu(\rv) = 0$ otherwise---with energy $E = E\sub{dimer} = -L^3$.

\begin{figure}
\putinscaledfigure{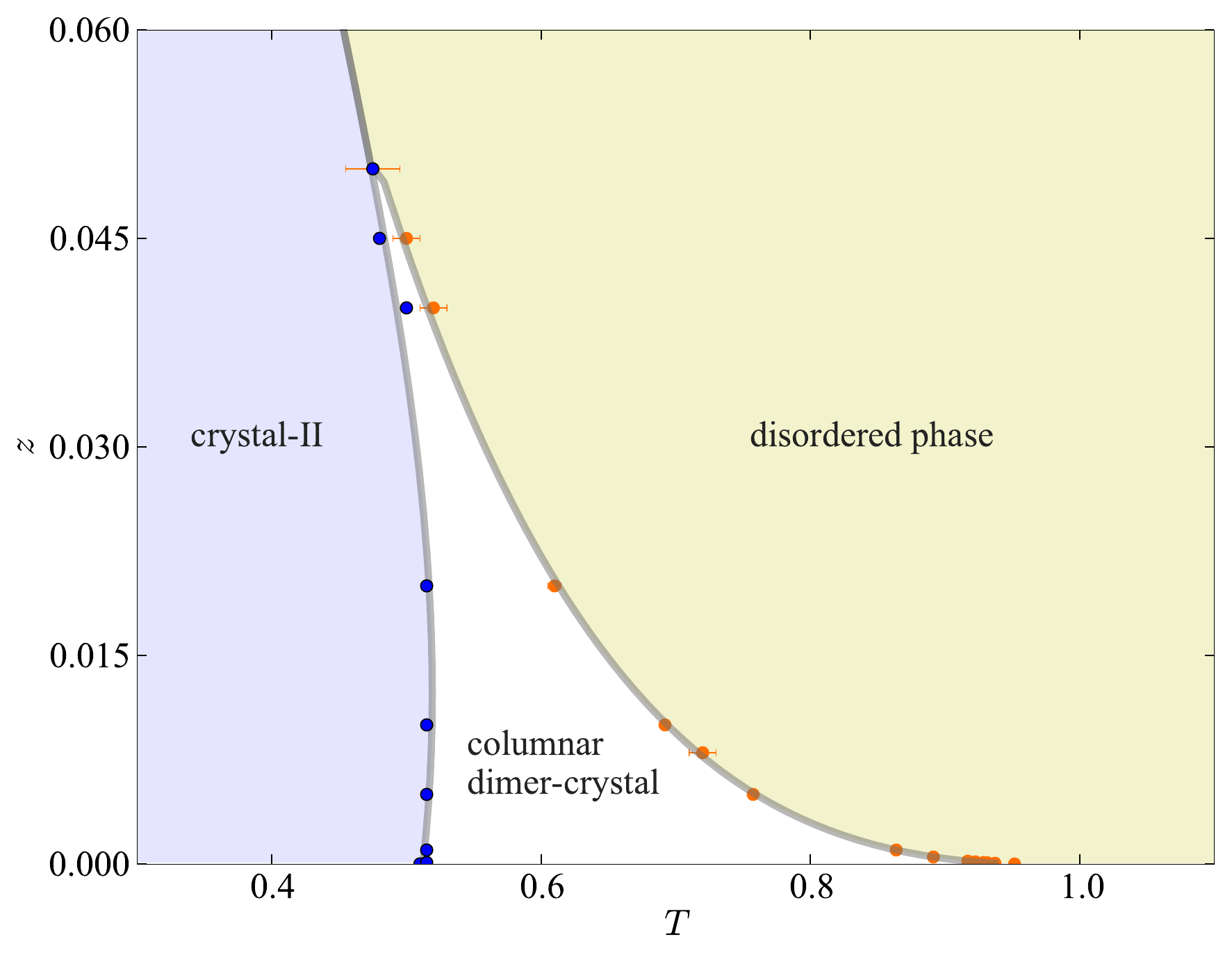}
\caption{Phase diagram for the cubic dimer model of \refeq{EqConfigEnergy}, with $v_2 = -1$ and $v_4 = +1$, as a function of temperature $T$ and monomer fugacity $z = \ee^{-\Delta/T}$. Points show results of Monte Carlo simulations, while the lines are guides to the eye. The columnar dimer crystal and thermally disordered phase are illustrated in \reffig{FigPhases}(a) and (c) respectively, while the Coulomb phase, \reffig{FigPhases}(b), is the limit $z\rightarrow 0$ of the disordered phase. The phase labeled Crystal II\cite{Charrier2} has a more complicated ordered structure, and is not of direct interest here. For $z\ge0.05$, we can resolve only a single first-order transition directly into this phase as temperature is reduced.
\label{FigPhaseDiagram}}
\end{figure}

The broken symmetry in these configurations is characterized by the ``magnetization'' (per site),
\beq{EqDefinem}
m_\mu = \frac{2}{L^3}\sum_{\rv} (-1)^{r_\mu} d_\mu(\rv)\punc{.}
\eeq
The $6$ minimal-energy configurations maximize $\lvert \mv \rvert$, having $\mv \in \{\pm\deltav_x,\pm\deltav_y,\pm\deltav_z\}$. The order parameter $\lvert\langle \mv \rangle\rvert$ is reduced by fluctuations for $T>0$ but it remains nonzero until a critical temperature $T\sub{c}$, at which point it vanishes and the system enters a phase without order. This phase boundary, between a columnar dimer crystal and a disordered phase, is illustrated in \reffig{FigPhaseDiagram} (which shows results of Monte Carlo simulations; see \refsec{SecResults}).

This description of the phase structure, based on the order parameter, is essentially identical whether $\Delta$ is infinite or merely large. (Decreasing $\Delta$ from infinity has the quantitative effect of reducing the order--disorder transition temperature $T\sub{c}$, because it favors disorder in the dimer degrees of freedom.) From the point of view of the Landau criterion for phases and transitions, there is no distinction between these two cases.

In fact, the two phases at $T > T\sub{c}$, illustrated in \reffig{FigPhases}(b) and (c) and occurring when $\Delta$ is, respectively, infinite and finite, are qualitatively distinct. As we will argue, and as our numerical results explicitly demonstrate, there is an important distinction between the transitions in these two cases, based on topological order and deconfinement. To clarify this distinction, we review in \refsec{SecGaugeTheory} the gauge-theoretical description of the dimer model.\cite{Huse}

The three phases of interest are not qualitatively affected by nonzero $v_4$, although for large positive $v_4$ the columnar states no longer minimize the energy and an additional ordered phase (``Crystal II'') can appear at low $T$.\cite{Charrier2} More importantly, the transition from the columnar dimer crystal into the higher-temperature phases changes from continuous to first order as $v_4$ is decreased below a value $v_{4\text{c}} \lesssim 0$.\cite{Charrier2} The tricritical point separating these two types of transition apparently strongly influences the effective critical properties near $v_4 = 0$.\cite{Alet,Charrier2,Papanikolaou} In our numerical results, we therefore use the value $v_4 = +1$, which allows us to access the critical behavior of the continuous confinement transition, while being small enough to avoid complications from additional ordered phases (see \reffig{FigPhaseDiagram} and \refsec{SecResultsPhaseStructure}).

\subsection{Effective gauge theory}
\label{SecGaugeTheory}

In the limit $\Delta = \infty$, the dimer model is constrained to have one dimer per site. This constraint can be rewritten in the form of a Gauss law by defining a ``magnetic field''\cite{Huse,HenleyReview}
\beq{EqDefineB}
B_\mu(\rv) = \eta_{\rv}\left[d_\mu(\rv) - \frac{1}{\goz}\right]\punc{,}
\eeq
where $\goz = 6$ is the coordination number, and $\eta_{\rv} = (-1)^{\sum_\mu r_\mu}$ is $\pm 1$ on the two sublattices. The lattice divergence, defined by
\beq{EqDefineDiv}
\Div_{\rv} B \equiv \sum_{\mu} [ B_\mu(\rv) - B_\mu(\rv - \deltav_\mu)]\punc{,}
\eeq
obeys
\beq{EqDivB}
\Div_{\rv} B = \eta_{\rv}[n(\rv) - 1]
\punc{,}
\eeq
and so monomers act as charges (``magnetic monopoles''), with sign depending on the sublattice.

In these terms, the partition function can be rewritten as
\beq{EqPartitionFunctionB}
\scZ = \sum_{\substack{\{B_\mu(\rv)\}\\ \lvert\Div_{\rv} B\rvert \le 1}} \ee^{-E\sub{dimer}/T} \prod_{\rv} z^{(\Div_{\rv} B)^2}\punc{,}
\eeq
where $z = \ee^{-\Delta/T}$ is the monomer fugacity. In particular, when $\Delta = \infty$, $z = 0$, and so
\beq{EqPartitionFunctionConstrained}
\scZ = \sum_{\substack{\{B_\mu(\rv)\}\\ \Div_{\rv} B = 0}} \ee^{-E\sub{dimer}/T}\punc{.}
\eeq

The configurations to which the system is constrained at $z=0$, viz.\ those of close-packed, hard-core dimers, satisfy a lattice Gauss law, $\Div_{\rv} B = 0$. Such a constraint is preserved by an appropriately chosen RG procedure, and is therefore expected to have important consequences for the long-distance physics, qualitatively distinguishing the cases with $z=0$ and $z>0$.

More explicitly, the result of such a procedure is to replace $B_\mu(\rv)$ by a coarse-grained continuum vector field $\Bv(\rv)$ with a corresponding constraint on its continuum divergence, $\del\cdot \Bv = 0$. In the Coulomb phase at $T > T\sub{c}$, where $\Bv(\rv)$ continues to fluctuate subject to this constraint, the effective continuum action density is
\beq{EqCoulombPhaseAction}
\scL\sub{gauge} = \frac{1}{2}K\lvert \Bv \rvert^2 = \frac{1}{2}K\lvert \del\times\Av \rvert^2\punc{,}
\eeq
where $\Bv = \del\times\Av$. This is a (Gaussian) fixed point under the RG, at which all other analytic terms are irrelevant, and at which correlations are algebraic.\cite{Huse,HenleyReview}

While higher-order terms are irrelevant, and a mass term for $\Av$ is forbidden by gauge invariance, monomer fugacity is a relevant perturbation at the Coulomb-phase fixed point. For any nonzero $z$, the algebraic correlations are cut off at a length scale $\propto z^{-1/d}$ set by the monomer separation.

\subsection{Confinement and deconfinement}

The magnetization $\mv$ is a local quantity that provides a conventional order parameter for the low-temperature phase. At $z=0$, the phase transition can alternatively be characterized through the concept of confinement.

Consider first the partition function of \refeq{EqPartitionFunctionConstrained}, but evaluated in the presence of a set of charges at fixed positions,
\beq{EqDefineZQ}
\scZ[Q_{\rv}] = \sum_{\substack{\{B_\mu(\rv)\}\\ \Div_{\rv} B = Q_{\rv}}} \ee^{-E\sub{dimer}/T}\punc{.}
\eeq
The ``monopole distribution function'' $G\sub{m}$ can then be defined, in terms of a test pair of oppositely charged monomers at $\rv_{\pm}$, as
\beq{EqDefineGm}
G\sub{m}(\rv_+,\rv_-) = \frac{\scZ[\delta_{\rv,\rv_+} - \delta_{\rv,\rv_-}]}{\scZ}.
\eeq
The function $G\sub{m}$ can be considered as the correlation function corresponding to the monomer fugacity.\cite{MonopoleScalingPRB} It is related to the effective interaction, $U\sub{m} = -T \ln G\sub{m}(\rv_+,\rv_-)$ induced between the monomers as a result of the fluctuating dimers.

To understand the behavior of $G\sub{m}$, imagine starting in a defect-free system and taking out a single dimer. This leaves two vacant sites, which are on opposite sublattices and hence carry opposite gauge charge. One can move each monomer by rearranging the dimers; all local rearrangements preserve the signs of the charges. In the ordered phase, attempting to separate the monomers in this way leaves behind a trail of disturbance, costing an energy (at least) proportional to its length, so $U\sub{m} \propto R = \lvert \rv_+ - \rv_-\rvert$. In the Coulomb phase, there is no order and so moving the monomers simply scrambles an already-disordered background, leaving no such trail. The scrambling induces only an entropic interaction; the monomers act like point charges in the effective magnetic field, and $U\sub{m}$ obeys the Coulomb law, $U\sub{m}\propto R^{-1}$.

For $z=0$, the large-separation limit of $G\sub{m}$ is therefore qualitatively distinct in the phases on either side of $T\sub{c}$. In the ordered phase at $T < T\sub{c}$, the monomers are confined, costing an unbounded energy to be separated to infinity, and so $G\sub{m} \rightarrow 0$. In the Coulomb phase at $T > T\sub{c}$, $U\sub{m}$ has a finite limit, and so they are deconfined, with a nonzero limit for $G\sub{m}$.

This distinction applies only for $z = 0$; for nonzero monomer fugacity, charges are screened in the limit of large separation, and $U\sub{m}$ always approaches a finite constant for large separation.

\section{Phase transitions}
\label{SecPhaseTransitions}

\subsection{Critical theories}
\label{SecCriticalTheories}

We are interested in the phase transition from the dimer crystal, with order parameter $\langle\mv\rangle \neq \zerov$, into phases in which the magnetization vanishes.

At $z=0$, symmetry-breaking order appears simultaneously with the confinement of monomers, and both of these effects must be incorporated in a critical theory. It has previously been argued\cite{CubicDimers,Charrier,Chen} that the transition is described by a Higgs theory of a noncompact $\mathrm{U}(1)$ gauge field $\Av$ and $\mathrm{SU(2)}$-symmetric matter fields $\vphiv$. The action density is
\beq{EqCriticalAction}
\begin{aligned}
\scL\sub{critical} &= \scL\sub{gauge} + \scL\sub{matter} + \scL\sub{matter--gauge}\\
\scL\sub{matter} &= s\lvert\vphiv\rvert^2 + u(\lvert\vphiv\rvert^2)^2\\
\scL\sub{matter--gauge} &= \lvert(\del - \ii \Av) \vphiv\rvert^2\punc{;}
\end{aligned}
\eeq
the pure-gauge part of the action $\scL\sub{gauge}$ is as given in \refeq{EqCoulombPhaseAction}. Note that the matter field $\vphiv$ is minimally coupled to $\Av$ and so, in the language where $\Bv = \del \times \Av$ is the magnetic field, has electric charge.\cite{FootnoteDualLanguage}

The transition occurs when $s$ is tuned through its critical value $s\sub{c}$. For $s < s\sub{c}$, corresponding to the lower-temperature phase, $\vphiv$ condenses and $\Av$ acquires an effective mass term by the Anderson--Higgs mechanism. This in turn eliminates the algebraic correlations and confines the monomers (magnetic charges) through the Meissner effect. Symmetry arguments show that the magnetization order parameter obeys $m_\mu \sim \vphiv^\dagger \sigmam^\mu \vphiv$, where $\sigmam^\mu$ is a Pauli matrix, and so becomes nonzero when $\vphiv$ condenses. The continuous $\mathrm{SU}(2)$ symmetry of $\scL\sub{critical}$ is broken by an eighth-order (in $\vphiv$) term,
\beq{EqCubicAnisotropy}
\scL\sub{cubic} = -u\sub{cubic} \sum_\mu m_\mu^4\punc{,}
\eeq
which is irrelevant at the critical point, but for $s < s\sub{c}$ selects states where $\langle \mv \rangle$ is aligned with one of the 6 cubic directions ($u\sub{cubic} > 0$).

For $z>0$, we instead expect the transition, between the ordered dimer crystal and the thermally disordered phase, to be described by a conventional Landau theory. The order parameter is the $3$-component vector $\langle \mv \rangle$, which preferentially aligns with one of the six cubic directions. The Landau action is therefore given by
\beq{EqLandauAction}
\scL\sub{Landau} = \lvert \del \mv \rvert^2 + s_{\mv} \lvert \mv \rvert^2 + u_{\mv} (\lvert \mv \rvert^2)^2 + \scL\sub{cubic}\punc{,}
\eeq
where $\scL\sub{cubic}$ is the cubic-anisotropy term given in \refeq{EqCubicAnisotropy}, again with $u\sub{cubic} > 0$.\cite{Footnotedivm} This theory can be viewed as the result of adding a nonzero density of magnetic monopoles to \refeq{EqCriticalAction}, suppressing the algebraic correlations of the gauge field $\Av$ and confining the matter field $\vphiv$ into the charge-neutral bilinear $\mv$.

In this case, the anisotropy $\scL\sub{cubic}$ is only quartic in the critical field $\mv$, and is known to be relevant. For $u\sub{cubic} = 0$, the transition would be continuous and in the Heisenberg universality class, but $u\sub{cubic} > 0$ is expected to drive it first order.\cite{Carmona} (In fact, this transition appears to be at most very weakly first-order; see \refsecand{SecScalingOD}{SecResultsPhaseStructure}.)

\subsection{Scaling}
\label{SecScaling}

The presence of the continuous confinement transition at $z=0$ and $T = T\sub{c}(z=0)$ implies that the behavior in its neighborhood is governed by the properties of the corresponding RG fixed point. This includes along the transition line at $T = T\sub{c}(z>0)$, and in particular determines the shape of the phase boundary as a function of $z$. In addition, the first-order nature of the transition at $z>0$ is apparently weak enough that, as we argue in \refsec{SecScalingOD}, scaling theory can be applied to this transition. These observations lead to a number of quantitative predictions that we are able to test in our simulations.

\subsubsection{Scaling near confinement transition}
\label{SecScalingConfinement}

Previous work\cite{MonopoleScalingPRL,MonopoleScalingPRB} has shown that, in addition to the reduced temperature
\beq{EqDefinet}
t = \frac{T - T\sub{c}(0)}{T\sub{c}(0)}\punc{,}
\eeq
there is a relevant scaling field at the confinement-transition fixed point given by the monomer fugacity $z$. The reduced free-energy density $f = -L^{-d}\ln \scZ$ therefore has a singular part obeying\cite{Cardy}
\beq{EqScalingfs}
f\sub{s}(t,z,h,L) \approx \lvert t \rvert^{2-\alpha}\Phi_{\pm}(z/\lvert t \rvert^\phi,L\lvert t \rvert^\nu,h/\lvert t \rvert^{2-\alpha-\beta})\punc{,}
\eeq
where $\Phi_{\pm}$ is a universal function, with the subscript $\pm$ indicating dependence on the sign of $t$. For completeness an applied field $h$ coupling to the magnetization has been included; unless stated otherwise we set $h=0$ in the following. The exponents $\alpha$, $\beta$, $\nu$, and $\phi$ are related to the RG eigenvalues $y_t$, $y_z$, and $y_h$ by
\begin{align}
\alpha &= 2 - \frac{d}{y_t}\\
\beta &= \frac{d - y_h}{y_t}\\
\nu &= \frac{1}{y_t}\label{EqDefinenu}\\
\phi &= \frac{y_z}{y_t}\punc{.}
\end{align}

\subsubsection{Phase boundary}
\label{SecScalingPhaseBoundary}

The phase boundary at $T = T\sub{c}(z)$ is a nonanalyticity of the free energy (for $L=\infty$), and hence of $\Phi_{-}$ [since $T\sub{c}(z) < T\sub{c}(0)$]. According to \refeq{EqScalingfs}, this boundary has the form
\beq{EqScalingTcz}
\lvert T\sub{c}(z) - T\sub{c}(0)\rvert \sim z^{1/\phi}\punc{.}
\eeq

While this result is asymptotically exact, there are substantial corrections for nonzero $z$ that must be incorporated for a correct interpretation of the numerical data. The most important comes from the leading irrelevant scaling variable at the fixed point, which we denote $u$, with RG eigenvalue $y_u = -\omega < 0$. Incorporating the unknown, nonuniversal constant $u$ into \refeq{EqScalingfs}, while setting $L=\infty$ for simplicity, we can write
\beq{EqScalingfs2}
f\sub{s}(t,z) \approx \lvert t \rvert^{2-\alpha}\Phi_{\pm}(z/\lvert t \rvert^\phi,u z^{\omega \phi/\nu})\punc{.}
\eeq
The value $\zeta\sub{c}$ of $\zeta = z/\lvert t \rvert^\phi$ at which $\Phi_-$ has a nonanalyticity is now dependent on $u z^{\omega \phi/\nu}$. Assuming that a Taylor expansion for $\zeta\sub{c}$ exists around $u = 0$ and dropping corrections of higher order in $z^{\omega\phi/\nu}$ gives
\beq{EqScalingTcz2}
\lvert T\sub{c}(z) - T\sub{c}(0)\rvert \sim z^{1/\phi}(1 + C z^{\omega\phi/\nu})\punc{,}
\eeq
where $C$ is a nonuniversal constant.

A functional-RG study\cite{Bartosch} of the Higgs theory, \refeq{EqCriticalAction}, found a small value of the correction exponent, $\omega \simeq 0.3$. This implies that the second term in \refeq{EqScalingTcz2}, as well as omitted higher-order corrections, should be significant, and our results are qualitatively consistent with this. It has in fact been argued that there may be logarithmic corrections to scaling in the JQ model,\cite{Sandvik,Pujari} which would imply that $\omega = 0$ at this fixed point; we are unable to exclude this possibility.

\subsubsection{Scaling of monopole distribution function}
\label{SecScalingMonopoleDistribution}

The monopole distribution function defined in \refeq{EqDefineGm} is the correlation function corresponding to the scaling field $z$, and so for $z=0$ has scaling form
\beq{EqGmScaling}
G\sub{m}(R; T, L) \approx R^{-2(d-y_z)} \Gamma_{\text{m}\pm}(R\lvert t\rvert^\nu,R/L)\punc{,}
\eeq
where $R = \lvert \rv_+ - \rv_- \rvert$ is the separation\cite{FootnoteIsotropic} and $\Gamma_{\text{m}\pm}$ is a universal function. This expression allows one to determine the RG eigenvalue $y_z$, and hence the exponent $\phi$, by measuring the monopole correlation function at the critical point, $t=0$ and $z=0$. Together with \refeq{EqScalingTcz}, this provides a quantitative prediction for the phase boundary at $z>0$.

\subsubsection{Scaling near order--disorder transition}
\label{SecScalingOD}

The order--disorder transition at $z>0$ is described by the Landau theory of \refeq{EqLandauAction}, an $\mathrm{O}(3)$ vector model with cubic anisotropy, $u\sub{cubic} > 0$. At the Heisenberg-class fixed point, $u\sub{cubic}$ is relevant with RG eigenvalue $\primecubic{y} > 0$, causing flow to a discontinuity fixed point, describing a first-order transition.\cite{Carmona,Cardy}

An argument based on RG theory, however, implies that, for small $z$, this first-order behavior requires inaccessibly large $L$. For small $z>0$ and $T\simeq T\sub{c}$, the trajectory of the RG flow is initially through the neighborhood of the confinement-transition fixed point. The cubic anistropy is irrelevant at this point, with RG eigenvalue $\noprimecubic{y} < 0$, and is consequently renormalized to a smaller value $\primecubic{u} \sim \noprimecubic{u} \ee^{\ell \noprimecubic{y}}$, where $\ell$, the ``RG time'' spent in this part of the trajectory, depends on $z$ as $\ee^{\ell} \sim z^{-1/y_z}$. The RG flow subsequently passes through the neighborhood of the Heisenberg fixed point, requiring time $\ee^{\ell'}\sim (\primecubic{u})^{1/\primecubic{y}}$ to reach the outflow trajectory leading to the discontinuity fixed point. The required length scale to reach the latter fixed point, and hence to see a first-order transition, is therefore $\sim \ee^{\ell}\ee^{\ell'}\sim z^{-\frac{1}{y_z}(1-\noprimecubic{y}/\primecubic{y})}$. (This additional length scale is a consquence of the dangerous irrelevant scaling variable $\noprimecubic{u}$ at the confinement fixed point. The same result can equivalently be derived using only scaling theory, following similar logic to that in \refsec{SecBinderSlope}.)

Cubic anistropy is apparently only very weakly relevant at the Heisenberg fixed point, with $\primecubic{y} \simeq 0.01$,\cite{Carmona} and so the length scale grows extremely rapidly with decreasing $z$. In our numerical results (see \refsec{SecResultsPhaseStructure}), we indeed find no clear signatures of a first-order transition at small $z$, and instead see results consistent with scaling up to the largest accessible system sizes.

On moderate length scales, one expects behavior governed by the Heisenberg fixed point, at which the relevant scaling fields are the reduced temperature $t' \propto T - T\sub{c}(z)$, defined with respect to $T\sub{c}(z)$, and the applied field $h$. Since this transition belongs in a different universality class from the confinement transition, the corresponding RG eigenvalues and critical exponents are different. We define $\nu'$($\neq \nu$) as the correlation-length exponent at the order--disorder transition.

Near the order--disorder fixed point, the fugacity $z$ determines only nonuniversal quantities, such as the values of irrelevant scaling fields. As $z$ is reduced, one expects a crossover to behavior described by the confinement transition, and hence a breakdown of the order--disorder criticality. On general grounds, this should occur when the characteristic length scale for monomers,\cite{MonopoleScalingPRB} $\lambda\sub{m} \sim z^{-\nu/\phi}$, exceeds the system size.

\subsubsection{Binder cumulant}

The magnetization $\langle\mv\rangle$ provides an order parameter for symmetry breaking at both the confinement transition ($z=0$) and the order--disorder transition ($z>0$). The corresponding Binder cumulant,\cite{Binder}
\beq{EqDefineBinder}
\scB = 1 - \frac{\langle \lvert\mv\rvert^4 \rangle}{3\langle \lvert\mv\rvert^2\rangle^2}\punc{,}
\eeq
therefore provides a particularly useful quantity for studying the crossover between the two critical behaviors.

The scaling dimension of $\scB$ vanishes at the fixed points corresponding to both transition classes, giving
\begin{align}
\scB &\approx \Phi_{\scB}(t L^{1/\nu},z/\lvert t \rvert^\phi) && \text{(confinement)}\label{EqBinderScaling}\\
\scB &\approx \Psi_{\scB}\blp g(z) [T-T\sub{c}(z)] L^{1/\nu'}\brp && \text{(order--disorder),}\label{EqBinderScalingOD0}
\end{align}
where the function $g(z)$ incorporates the dependence on $z$ of the nonuniversal constant of proportionality. Near the transition for small nonzero $z$, both \refeqand{EqBinderScaling}{EqBinderScalingOD0} apply, and the requirement that they be consistent fixes
\beq{EqScalinggz}
g(z) \sim z^{(\nu - \nu')/(\phi \nu')}\punc{.}
\eeq

\subsubsection{Binder cumulant crossings}

According to \refeqand{EqBinderScaling}{EqBinderScalingOD0}, $\scB$ is independent of $L$ at the transition, implying that plots of the Binder cumulant versus temperature for different $L$ cross at $T = T\sub{c}(z)$. Corrections to scaling in fact imply that the crossing points are weakly dependent on $L$. Including leading-order corrections at the order--disorder transition replaces \refeq{EqBinderScalingOD0} by
\begin{multline}
\label{EqBinderScalingOD1}
\scB \approx \Psi_{\scB}\blp g(z)[T-T\sub{c}(z)]L^{1/\nu'}\brp \\{}+ L^{-\omega'}u(z)\tilde\Psi_{\scB}\blp g(z)[T-T\sub{c}(z)]L^{1/\nu'}\brp\punc{,}
\end{multline}
where $\tilde\Psi_{\scB}$ is an unknown function, $-\omega'$ is the RG eigenvalue of the leading irrelevant scaling operator, and $u(z)$ expresses the dependence of its coefficient on $z$. Again requiring consistency with \refeq{EqBinderScaling}, we find
\beq{EqScalinguz}
u(z) \sim z^{-\omega' \nu/\phi}\punc{.}
\eeq

At this order, the Binder cumulants at $L=L_1$ and $L_2$ have a crossing at $T = T_\times(L_1,L_2)$ given by\cite{Binder,Ferrenberg}
\beq{EqBinderCrossingOD}
T_\times(L_1,L_2) - T\sub{c}(z) \sim \frac{u(z)}{g(z)} \frac{L_2^{-\omega'} - L_1^{-\omega'}}{L_1^{1/\nu'} - L_2^{1/\nu'}}\punc{.}
\eeq
In applying this result to numerical data, it is convenient to fix the ratio $\rho = L_2/L_1$; incorporating the $z$ dependence of $g(z)$ and $u(z)$ then gives
\beq{EqBinderCrossingOD2}
T_\times(L,\rho L) - T\sub{c}(z) \sim z^{-(1+\omega'\nu - \nu/\nu')/\phi} L^{-\omega'-1/\nu'}\punc{.}
\eeq
For the confinement transition, one similarly finds
\beq{EqBinderCrossingConfinement}
T_\times(L,\rho L) - T\sub{c}(0) \sim L^{-\omega-1/\nu} \quad\text{($z=0$)} \punc{.}
\eeq

\subsubsection{Slope of Binder cumulant at crossing}
\label{SecBinderSlope}

The derivative of the Binder cumulant, taken with respect to $T$ and evaluated at $T = T\sub{c}(z)$, can be used to determine the correlation-length exponents $\nu$ and $\nu'$. At $z=0$, one finds the scaling result
\beq{EqBinderSlope}
\left.\parder{\scB}{T}\right\rvert_{T = T\sub{c}(0)} \!\sim L^{1/\nu}\punc{.}
\eeq
Similarly, for $z>0$, the slope is
\beq{EqBinderSlopeOD}
\left.\parder{\scB}{T}\right\rvert_{T = T\sub{c}(z)}
\!\approx \frac{g(z)}{T\sub{c}(0)}\left[\Psi_{\scB}'(0)L^{1/\nu'} + u(z)\tilde\Psi_{\scB}'(0)L^{1/\nu' - \omega'}\right]\punc{,}
\eeq
where the correction term from \refeq{EqBinderScalingOD1} has been included. In both cases, the slope is proportional to $L^{1/\nu^{(\prime)}}$ in the limit of large $L$.

The corrections in \refeq{EqBinderSlopeOD} have magnitude proportional to $u(z)$ and therefore become more significant for smaller $z$. In agreement with the general considerations of \refsec{SecScalingOD}, breakdown of order--disorder scaling occurs for system size $L$ below $\lambda\sub{m} \sim z^{-\nu/\phi}$, when the ``correction'' becomes larger than the zeroth-order term.

\subsubsection{Flux stiffness}
\label{SecFluxStiffness}

An additional quantity of interest at the confinement transition is the flux stiffness,\cite{Alet,FootnoteFlux}
\beq{EqDefineK}
K^{-1} = \frac{1}{3L^2}\sum_{r_\perp,\mu}\langle [\phi_\mu(r_\perp)]^2 \rangle\punc{,}
\eeq
where
\beq{EqDefineFlux}
\phi_\mu(r_\perp) = \sum_{\substack{\rv\\ r_\mu = r_\perp}} B_\mu(\rv)
\eeq
is the net ``magnetic flux'' through a closed surface, spanning the periodic boundaries, normal to $\deltav_\mu$. The flux stiffness $K^{-1}$ vanishes in the thermodynamic limit for $z=0$ and $T < T\sub{c}$, while it approaches a constant in the Coulomb phase.

At the confinement phase transition, the quantity $LK^{-1}$ has zero scaling dimension,\cite{Alet} and hence shows a similar crossing point to the Binder cumulant and a scaling form
\beq{EqStiffnessScaling}
LK^{-1}(t,z,L) \approx \Phi_K(tL^{1/\nu},z/\lvert t\rvert^\phi)\punc{.}
\eeq
The scaling dimension of the flux stiffness is not similarly fixed at the order--disorder transition and so, in contrast to the case of the Binder cumulant, the crossing points of $LK^{-1}$ do not converge for $z>0$.

\section{Numerical results}
\label{SecResults}

We have studied the dimer model defined in \refsec{SecModel} using a Monte Carlo method based on the directed loop algorithm\cite{SandvikMoessner} but adapted to allow for violations of the dimer constraint. An update within the set of constrained dimer configurations can be performed by shifting dimers along a closed loop. To change the number of monomers, one shifts dimers along an open path, creating or annihilating monomers at each end. (This corresponds to adding a string of magnetic flux with a monopole at each end.) Details of the algorithm are given in the Appendix.

Simulations were performed on lattices with periodic boundary conditions and $L^3$ sites, with system sizes up to $L = 144$.

\subsection{Phase structure}
\label{SecResultsPhaseStructure}

The phase structure of the model at $z=0$ has previously been studied extensively using Monte Carlo simulations.\cite{Alet,Misguich,Charrier,Chen,Papanikolaou,Charrier2} For $v_4 = 0$, one finds only the Coulomb phase and the columnar dimer crystal, while for $v_4 > 0$ an additional crystalline phase, denoted Crystal II, occurs at low $T$.\cite{Charrier2} We are mainly interested in the phase boundary between the Coulomb and columnar phases, which is shifted by nonzero $v_4$ but qualitatively unaffected.

The phase structure at nonzero $z$ and $v_4 = +1$ is shown in \reffig{FigPhaseDiagram}. The transition temperature to the columnar dimer crystal is reduced as $z$ increases, while the transition into Crystal II is largely unchanged. At larger $z = z\sub{max}\simeq 0.5$, the two phase boundaries merge into a single first-order transition between the disordered phase and Crystal II. This sets an upper limit of $z\sub{max}$ on the values of $z$ at which the transition to the columnar dimer crystal can be studied.

There is strong evidence that the transition at $z=0$ is continuous for $v_4 > 0$ but first order for $v_4 < v_{4\text{c}} \lesssim 0$.\cite{Charrier2} As noted in \refsecand{SecCriticalTheories}{SecScalingOD}, we expect that the transition at $z>0$ is first order, but only very weakly so for small $z$.
\begin{figure}
\putinscaledfigure{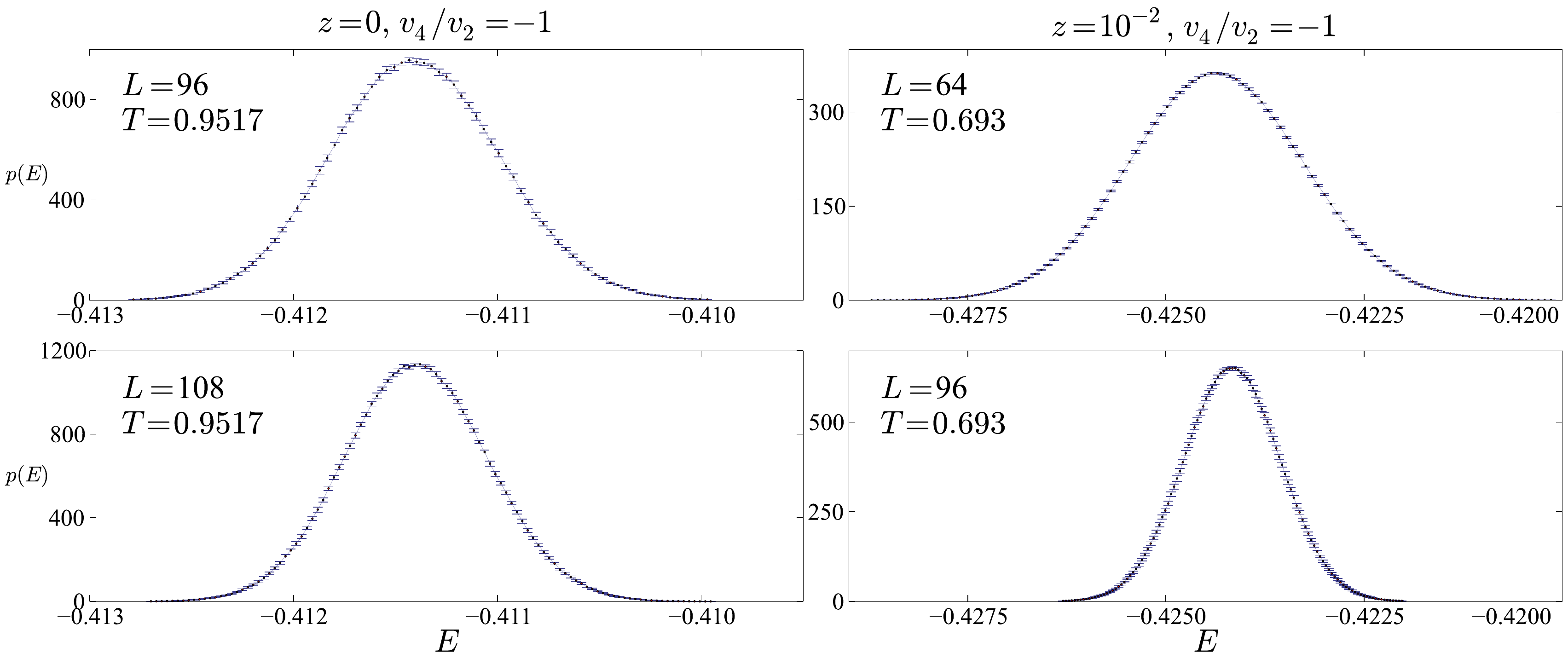}
\caption{Energy histograms from Monte Carlo simulations at $z=0$ (left) and $z=10^{-2}$ (right). There is no indication of a double-peak structure, which would indicate a first-order transition, within the accuracy obtained in the simulations.
\label{FigHistograms}}
\end{figure}
Energy histograms close to the critical temperature, shown in \reffig{FigHistograms}, exhibit no signs of the double-peak structure that would indicate a first-order transition. The latent heat, if nonzero, should grow with $z$, but only a single peak is resolvable (for our largest available system sizes) up to the maximum accessible $z = z\sub{max}$.

By contrast, for $v_4 = 0$, a first-order transition is clearly seen even for $z\simeq 10^{-2}$. In this case, the effects of the tricritical point at $z=0$ and $v_4 = v_{4\text{c}}$ are likely important\cite{Charrier2} and our scaling conclusions are less reliable.

\subsection{Confinement transition}
\label{SecResultsConfinement}

\begin{figure}
\putinscaledfigure{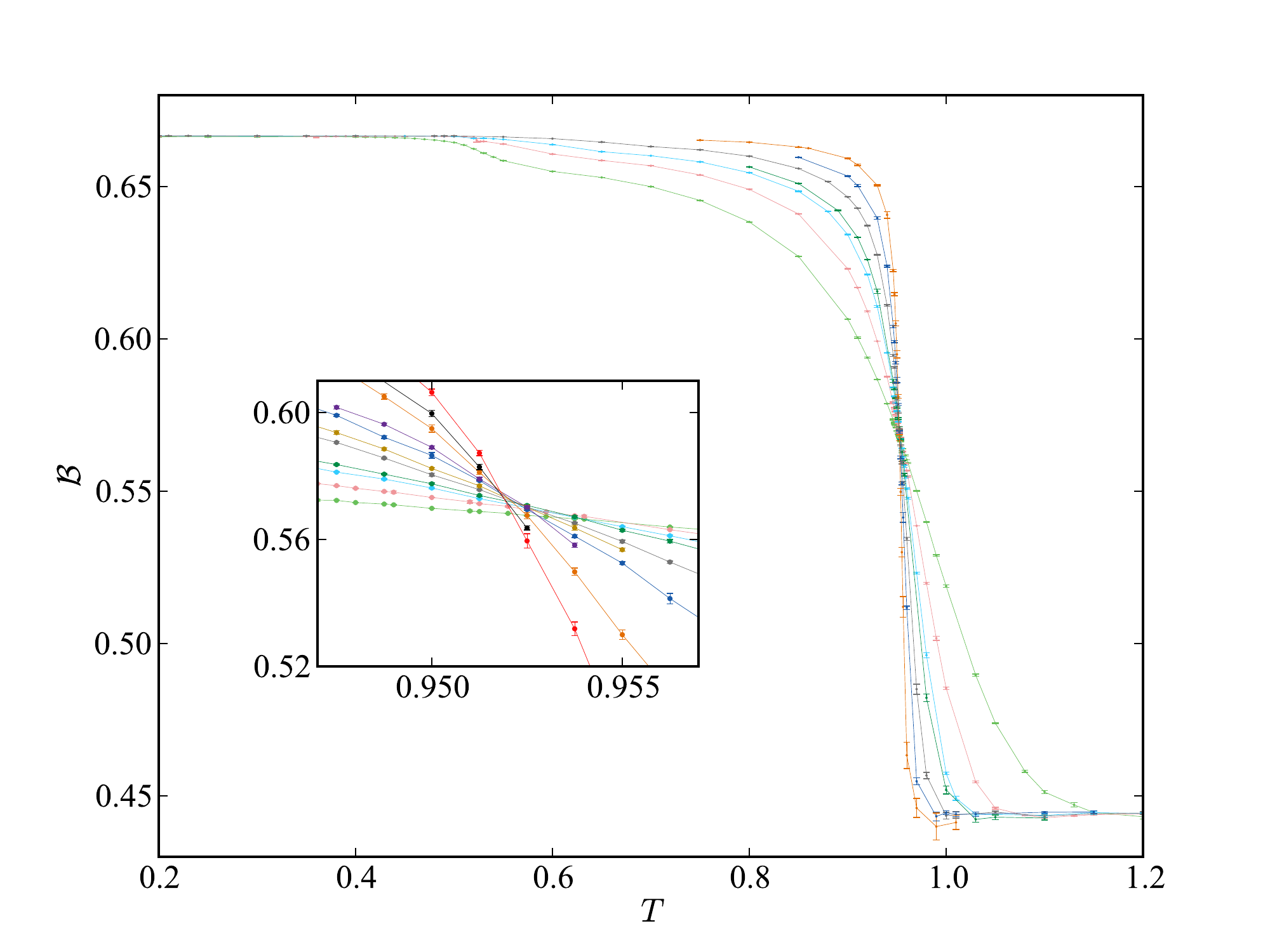}
\vspace{-0.8cm}
\putinscaledfigure{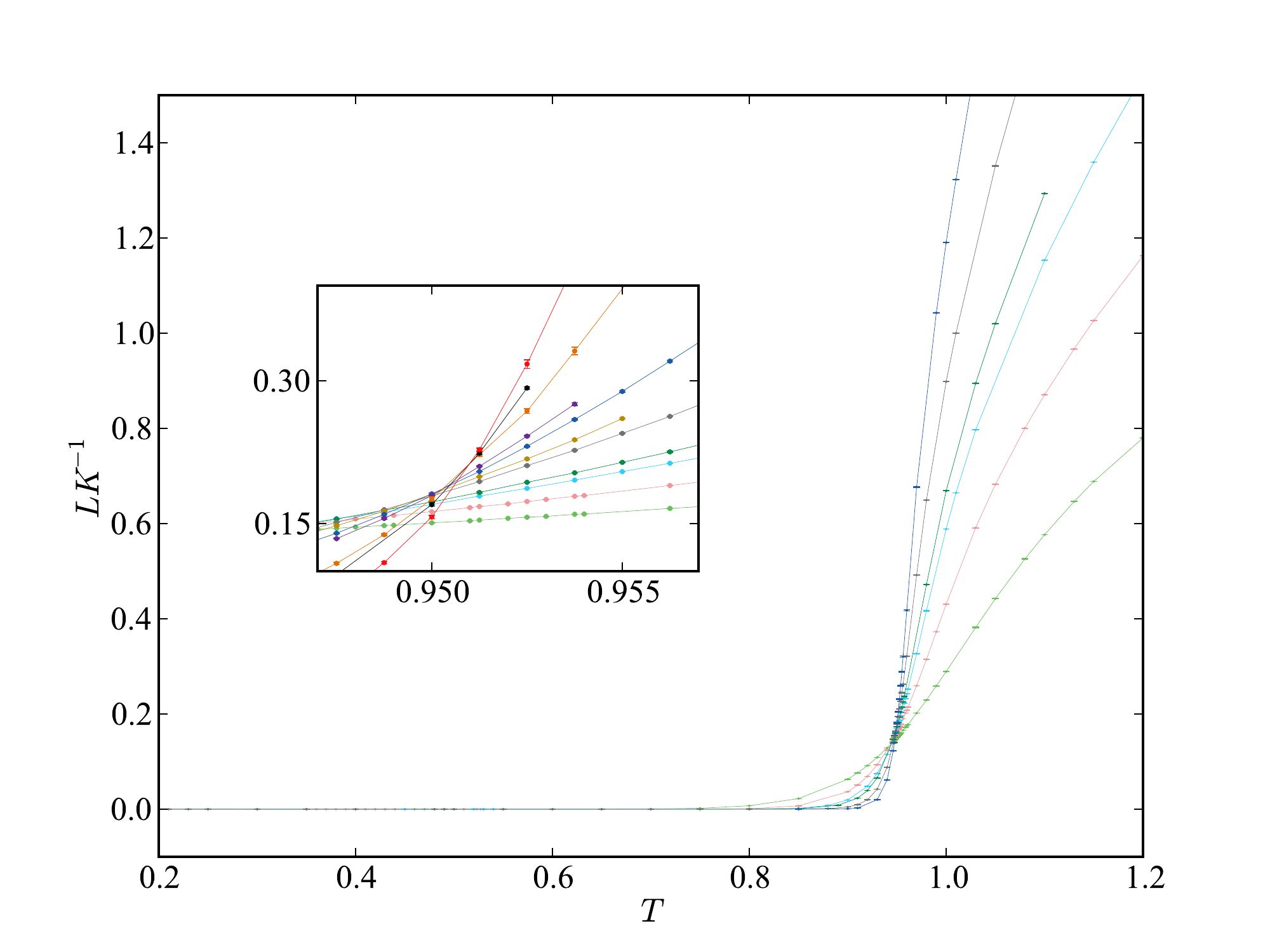}
\putinscaledfigure{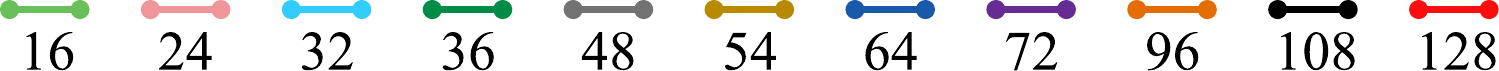}
\caption{Confinement transition at $z=0$ located using the Binder cumulant $\scB$ and flux stiffness $K^{-1}$. The quantities $\scB$ (top panel) and $LK^{-1}$ (bottom) are plotted versus temperature $T$ for various system sizes $L$. Both quantities show crossing points at the critical temperature in the limit of large $L$.
\label{FigBinderz0}}
\end{figure}

\begin{figure}
\putinscaledfigure{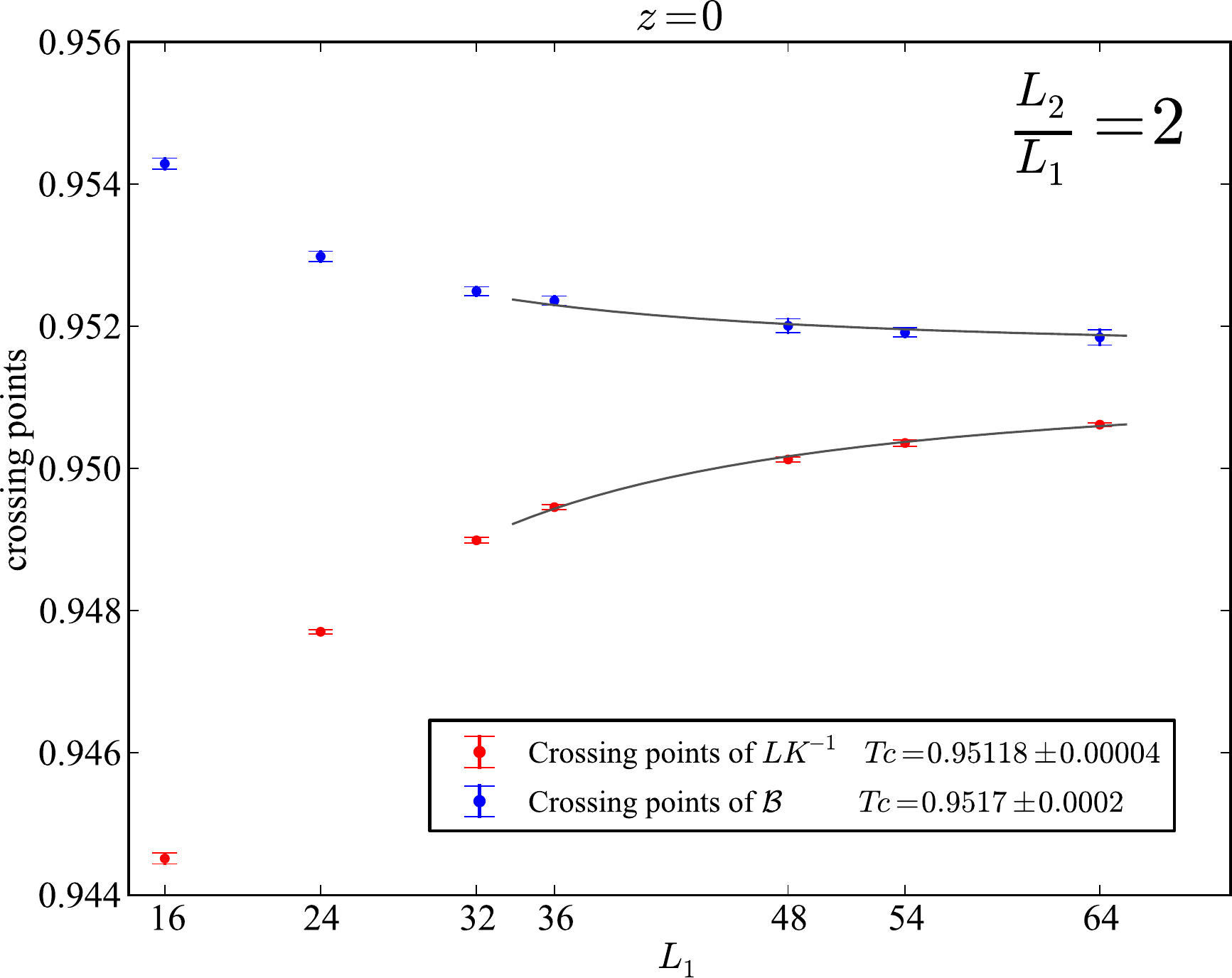}
\caption{Crossing points of $\scB$ (upper, blue points) and $LK^{-1}$ (lower, red points) at $z=0$, for pairs of systems sizes $L$ in the ratio $L_2/L_1 = 2$. The solid lines are fits to \refeq{EqBinderCrossingConfinement} and the corresponding expression for the stiffness, including points with $L_1\ge 36$, with only the critical temperature $T\sub{c}(z=0)$ and the constant of proportionality as fitting parameters. The exponent $-\omega -1/\nu = -1.9$ is fixed using $\nu=0.6$ (\refcite{Charrier2}) and $\omega=0.3$ (\refcite{Bartosch}). (The fitted critical temperature is not particularly sensitive to the value of the exponent; one can alternatively fit to this parameter, which gives a consistent value of $T\sub{c}$ but larger errors.)
\label{FigBinderCrossingsz0}}
\end{figure}

We begin by reporting our results for the transition at $z=0$ (and $v_4 = +1$), which has previously been studied in \refcite{Charrier2}. The Binder cumulant $\scB$ and flux stiffness $K^{-1}$ at $z=0$ are shown in \reffig{FigBinderz0} for various system sizes. The crossover between the low- and high-temperature values becomes steeper for larger $L$, with a sharply defined crossing point appearing.
The crossing temperatures for sizes $L$ in the ratio $L_2/L_1 = 2$ are shown in \reffig{FigBinderCrossingsz0}, along with fits to \refeq{EqBinderCrossingConfinement} and the corresponding expression for the stiffness.
These fits give values for the critical temperatures of $T\sub{c}(0) = 0.9517\pm 0.0002$ (Binder) and $T\sub{c}(0) = 0.95118 \pm 0.00004$ (stiffness).

The close agreement between the two methods of calculating $T\sub{c}(0)$ provide evidence that the ordering, measured by the Binder cumulant, and confinement, measured by flux stiffness, occur at a single transition. The remaining discrepancy, $\delta T\sub{c} = 0.0005 \pm 0.0002$, suggests either that our estimates of statistical errors are somewhat optimistic or that small systematic errors remain. (For each $L$, the crossing temperatures of $\scB$ and $LK^{-1}$ bound the true critical temperature.\cite{FootnoteBoundTc})

\begin{figure}
\putinscaledfigure{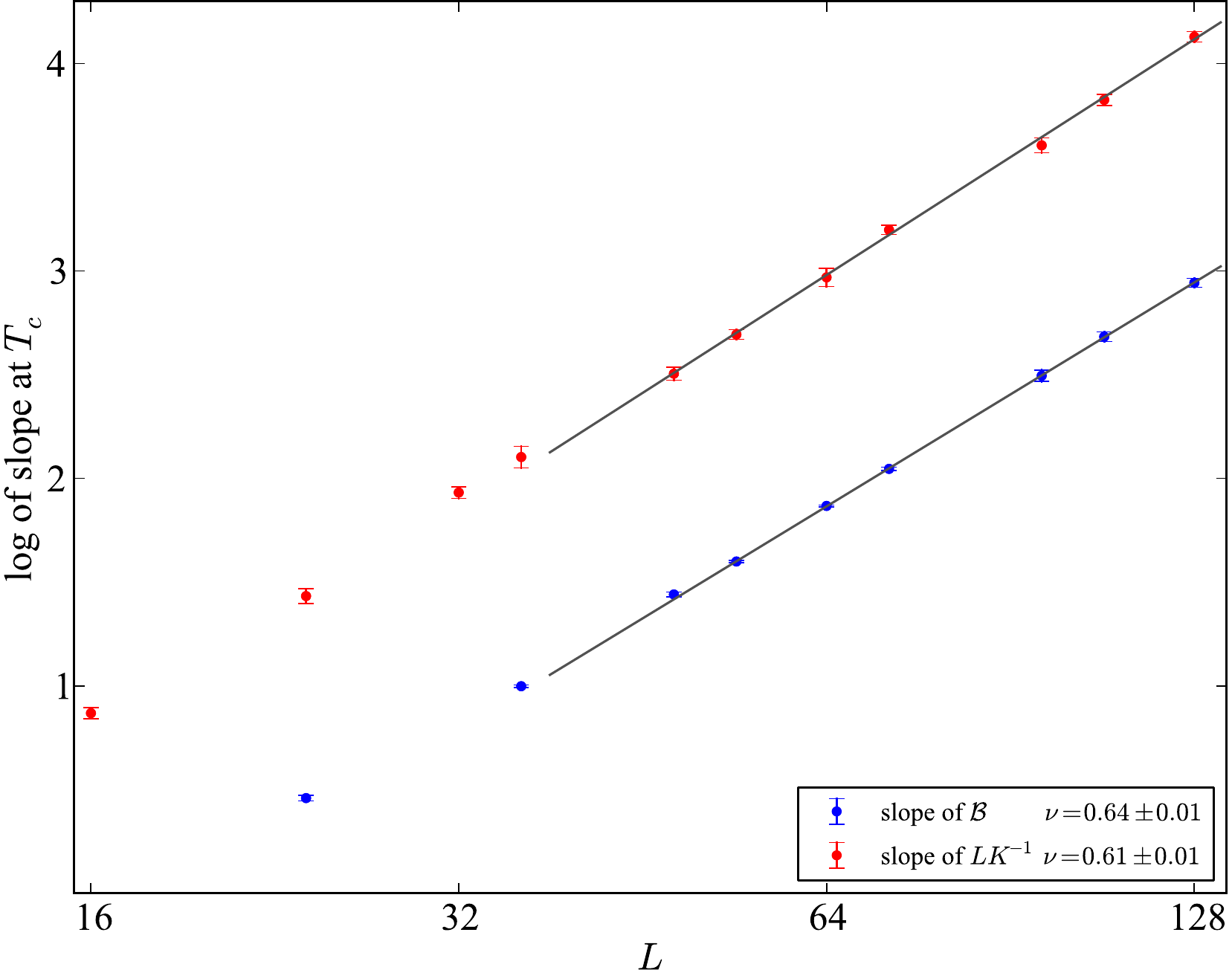}
\caption{Temperature-derivative of $\scB$ (lower, blue points) and of $L K^{-1}$ (upper, red points) evaluated at $T\sub{c}(z=0)$, versus system size $L$. Both are proportional, for large $L$, to $L^{1/\nu}$, giving values $\nu = 0.64 \pm 0.01$ (Binder cumulant) and $\nu = 0.61 \pm 0.01$ (stiffness). (In both cases, the fit includes only points with $L \ge 48$. The error bars include the uncertainty on $T\sub{c}$.)
\label{FigBinderSlopez0}}
\end{figure}

The value of $\nu$ can be determined from the derivative of the Binder cumulant at the critical temperature, as described in \refsec{SecBinderSlope}, or equivalently using the flux stiffness. The slope of $\scB$ and $LK^{-1}$ at $T\sub{c}(0)$ are shown in \reffig{FigBinderSlopez0}, as functions of $L$, along with fits to \refeq{EqBinderSlope} and a similar expression for the stiffness. We find $\nu = 0.64 \pm 0.01$ using the Binder cumulant and $\nu = 0.61 \pm 0.01$ using the stiffness.

Our results at $z=0$ are consistent with previous results in the CDM with $v_4 = +1$: Charrier and Alet\cite{Charrier2} found $T\sub{c} = 0.953 \pm 0.001$, $\nu = 0.60\pm0.04$ (Binder) and $\nu=0.61\pm0.04$ (stiffness).

\subsection{Monopole distribution function}
\label{SecResultsMonopoleDistribution}

As discussed in \refsec{SecScalingMonopoleDistribution}, the scaling behavior of the monopole distribution function $G\sub{m}(R=\lvert \rv_+ - \rv_-\rvert;T\sub{c},L)$ at $z=0$ can be used to determine $y_z = \phi/\nu$. To calculate $G\sub{m}$, defined in \refeq{EqDefineGm}, we require the partition function in the presence of a monopole--antimonopole pair, $\scZ[\delta_{\rv,\rv_+} - \delta_{\rv,\rv_-}]$. The directed-loop Monte Carlo algorithm that we use is well-suited to calculating this quantity, as the intermediate states appearing during the construction of a loop involve a pair of inserted monomers.\cite{SandvikMoessner}

\begin{figure}
\putinscaledfigure{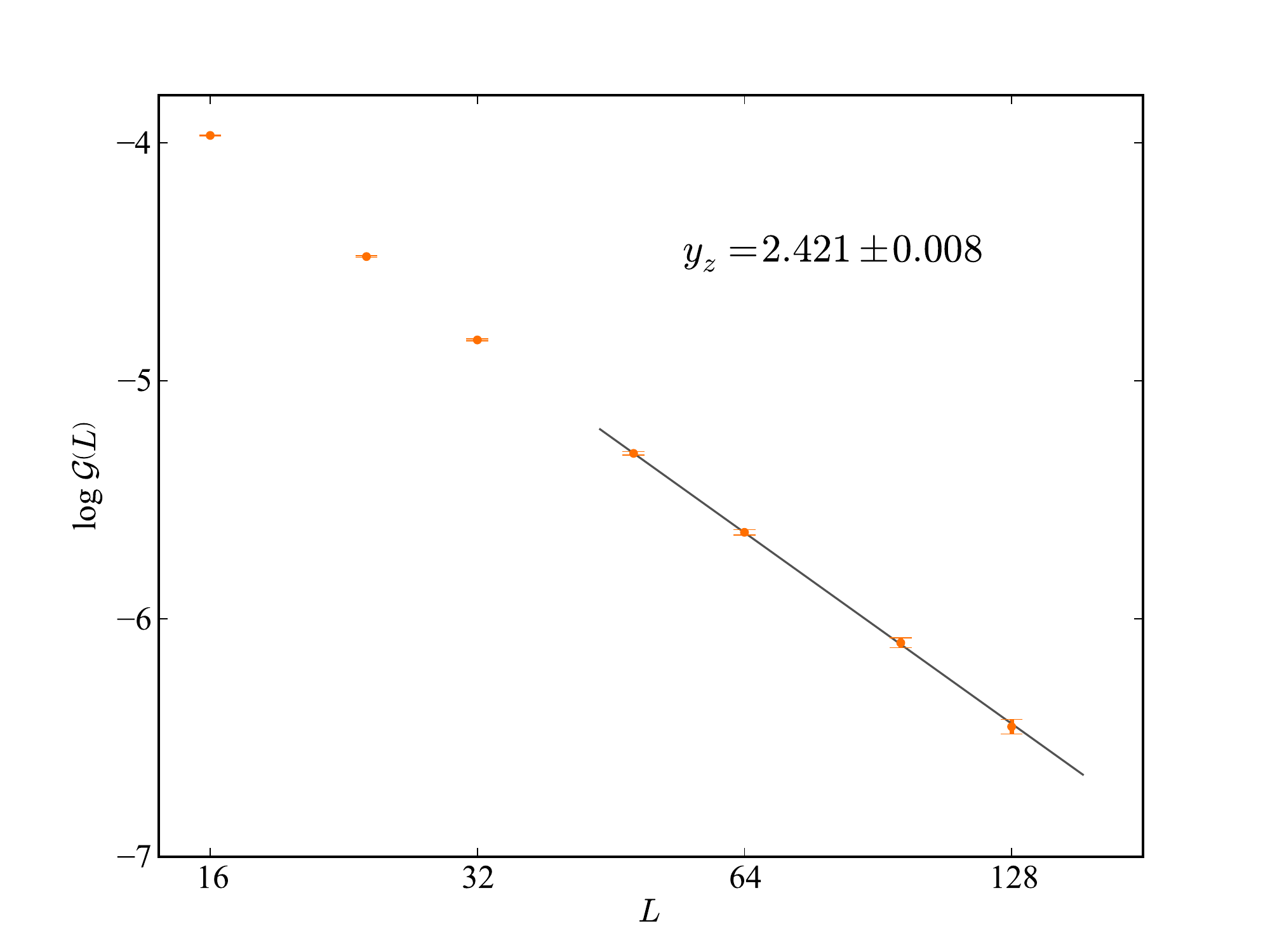}
\caption{Log--log plot of $\scG(L)$, the normalized value of the monopole distribution function $G\sub{m}$ (see \refsec{SecResultsMonopoleDistribution}), evaluated at the critical temperature. According to \refeq{EqGmScaling}, this ratio is asymptotically proportional to $L^{-2(d-y_z)}$; the line is a fit (to $L \ge 48$) with $y_z = 2.421\pm 0.008$.
\label{FigMonopoleDistribution}}
\end{figure}

Since the partition function can be calculated in Monte Carlo simulations only up to an $L$-dependent factor, we calculate the ratio
\beq{EqDefinescG}
\scG(L) = \frac{G\sub{m}(R\sub{max},L)}{G\sub{m}(R\sub{min},L)}\punc{,}
\eeq
with $R\sub{max}/L$ and $R\sub{min}$ both fixed and of order unity. This ratio, which is asymptotically proportional to $L^{-2(d-y_z)}$, is shown in \reffig{FigMonopoleDistribution}, along with a fit giving $y_z = 2.421 \pm 0.008$.
Using our estimate of $\nu$ from the Binder cumulant gives a crossover exponent $\phi = y_z \nu = 1.55 \pm 0.02$ (with the error dominated by that on $\nu$).

\subsection{Order--disorder transition}
\label{SecResultsODTransition}

\begin{figure}
\putinscaledfigure{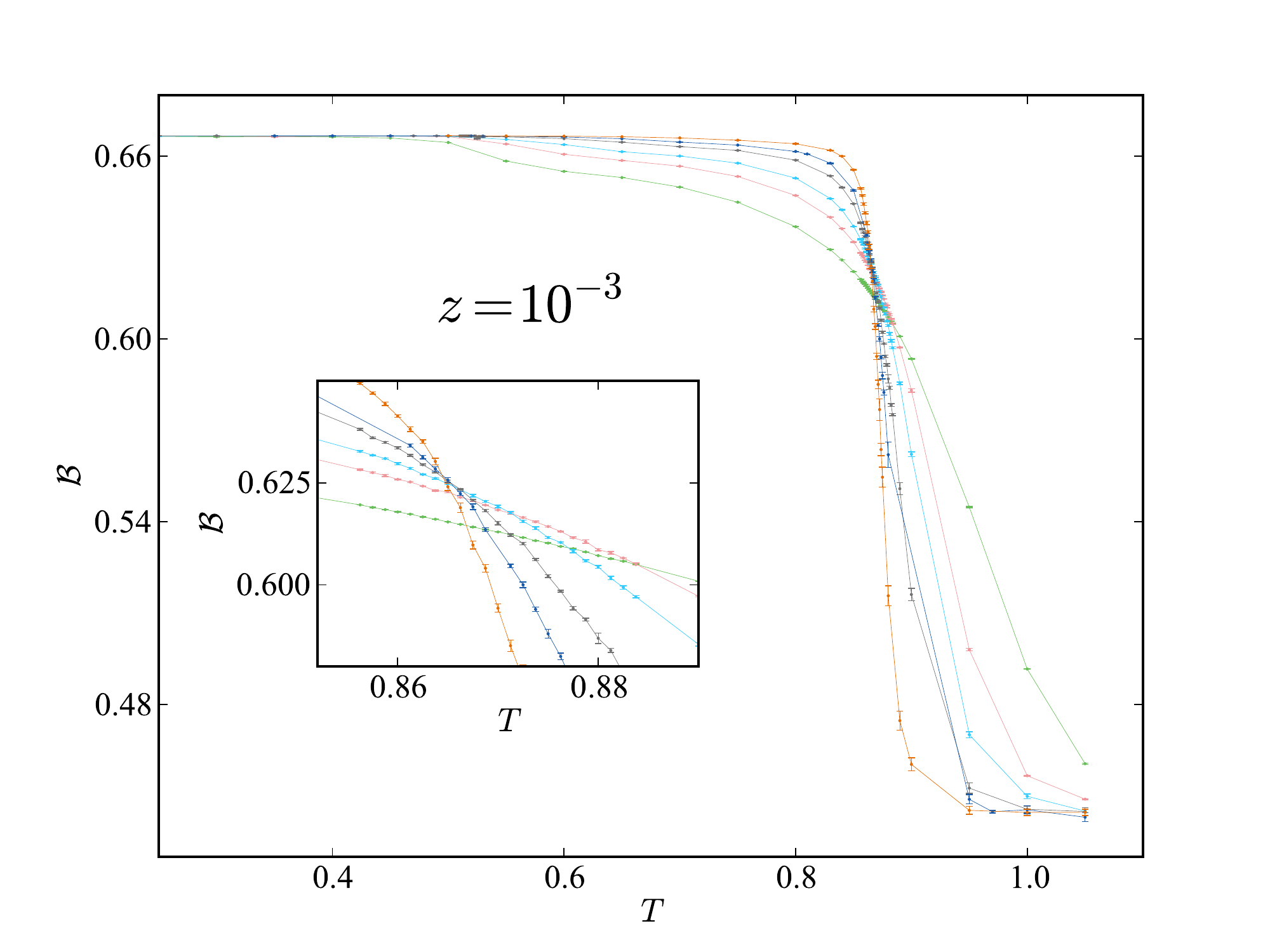}
\putinscaledfigure{legend}
\caption{Binder cumulant $\scB$ as a function of $T$ for $z=10^{-3}$. As for $z=0$ (\reffig{FigBinderz0}), curves for different $L$ cross at the critical temperature (for $L \gtrsim 48$), consistent with a continuous transition. 
\label{FigBinderOD1}}
\end{figure}

\reffigand{FigBinderOD1}{FigBinderOD2} show the Binder cumulant $\scB$ across the order--disorder transition for several values of $z$.
Consistent with an apparently continuous transition, the crossing points converge for the largest values of $L$; the convergence is slower (requiring larger $L$) for smaller $z$, in agreement with \refeq{EqBinderCrossingOD2}.
\begin{figure}
\includegraphics[width=0.7\columnwidth]{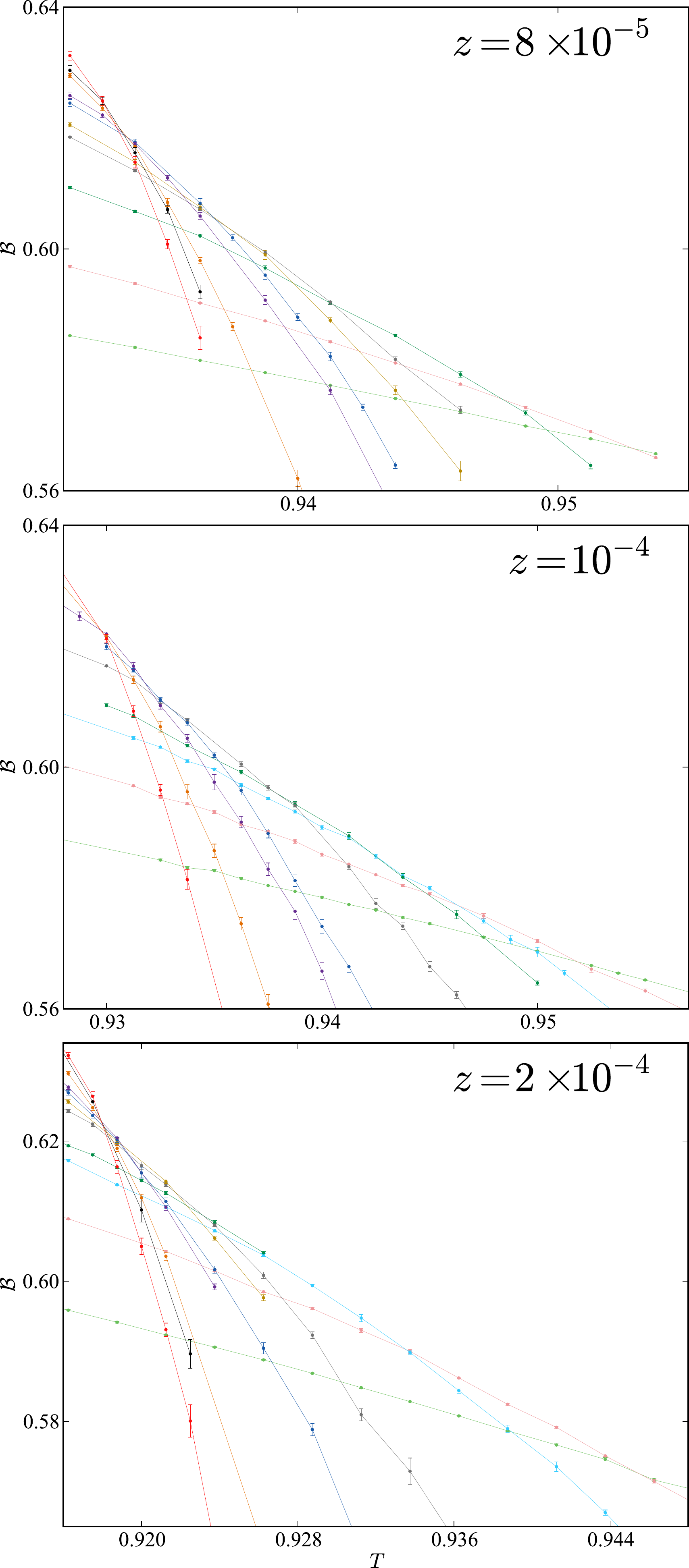}
\putinscaledfigure{legend}
\caption{Binder cumulant $\scB$ as a function of $T$ for several $z>0$. In each case, a well-defined crossing point develops for large $L$, but the convergence is slower for smaller $z$.
\label{FigBinderOD2}}
\end{figure}
For each value of $z$, polynomial interpolations are used to find the crossing points for pairs of system sizes $L$ in the ratio $L_2/L_1=2$. These are then fit to \refeq{EqBinderCrossingOD2} in order to find $T\sub{c}(z)$; examples are shown in \reffig{FigBinderCrossingsOD}.
\begin{figure}
\includegraphics[width=0.7\columnwidth]{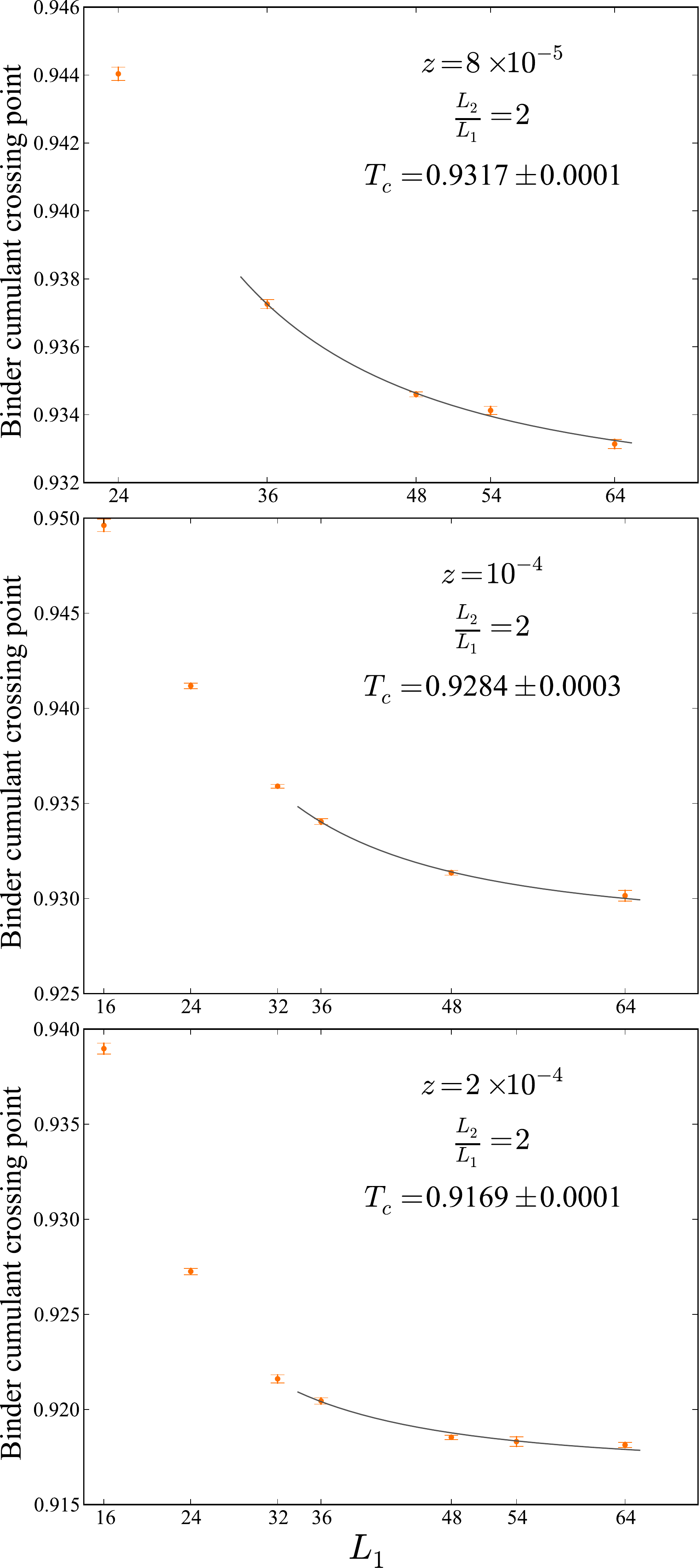}
\caption{Crossing points of Binder cumulants for several $z>0$ (shown in \reffig{FigBinderOD2}), for pairs of systems sizes $L$ in the ratio $L_2/L_1 = 2$. In each case, the solid line is a fit to \refeq{EqBinderCrossingOD2}, using fixed exponent $-\omega' - 1/\nu' = -2.2$, appropriate to the Heisenberg universality class.\cite{CampostriniHeisenberg} (Only sizes $L_1\ge 36$ are included in each fit.)
\label{FigBinderCrossingsOD}}
\end{figure}
Note that the fits use fixed values of $\omega'$ and $\nu'$ appropriate to the Heisenberg universality class.\cite{CampostriniHeisenberg} (The fitted values of $T\sub{c}$ are not particularly sensitive to these values; consistent results with larger error bars are found by fitting to the exponent.)

\begin{figure}
\putinscaledfigure{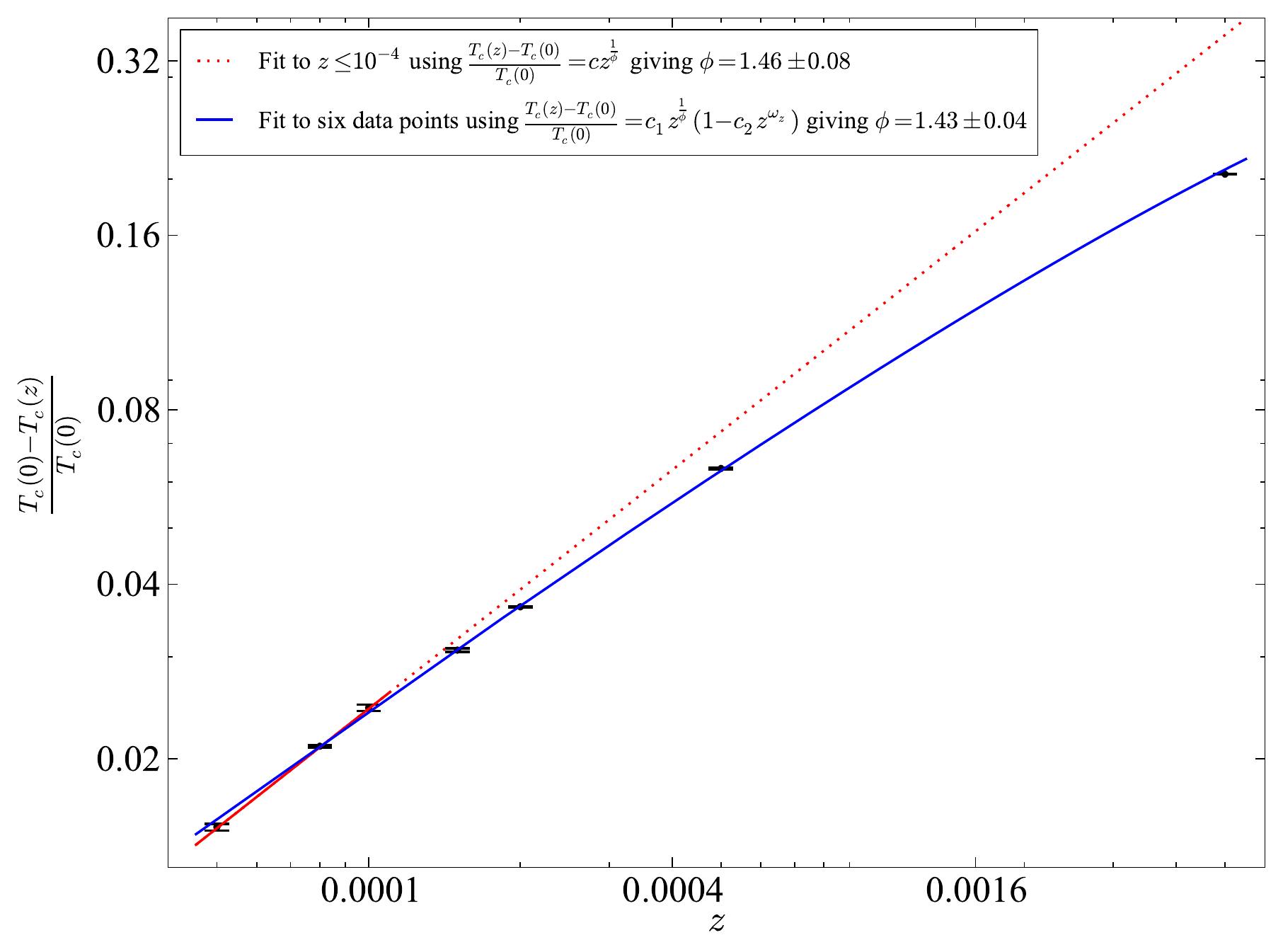}
\caption{Dependence of transition temperature $T\sub{c}$ on monomer fugacity $z$. For each $z$, $T\sub{c}(z)$ is determined as illustrated in \reffig{FigBinderCrossingsOD}. The red line is a fit to the scaling result $T\sub{c} \sim z^{1/\phi}$, using only the $3$ smallest values of $z$ ($z \le 10^{-4}$), giving $\phi = 1.46 \pm 0.08$. The blue curve fits $z \le 10^{-3}$ to \refeq{EqScalingTcz2}, which includes the leading corrections to scaling, giving $\phi = 1.43 \pm 0.05$.
\label{FigTcz}}
\end{figure}

The dependence of the transition temperature $T\sub{c}$ on monomer fugacity $z$ is shown in \reffig{FigTcz}.
We find that extremely small values of $z$ (on the order of $10^{-4}$) are required to give a power-law dependence of $T\sub{c}(z)$, indicating that the corrections to scaling are substantial. A power-law fit to the smallest three $z$ values, with $z \le 10^{-4}$, gives $\phi = 1.46\pm 0.08$, which is in good agreement with the value $\phi = 1.55 \pm 0.05$ found using calculations at $z=0$.

To incorporate leading-order corrections, we use \refeq{EqScalingTcz2}, with the value of the correction exponent $\omega=0.3$ found in a functional-RG study\cite{Bartosch} of the critical action, \refeq{EqCriticalAction}. This increases the range of $z$ over which a reasonable fit is found; we find a consistent value, $\phi = 1.43 \pm 0.04$, by including $z \le 10^{-3}$. (While the nominal error of the fit is reduced by including data at larger $z$, the quoted value is not necessarily reliable: The small magnitude of $\omega$ implies that significant higher-order corrections should also be expected.)

\subsection{Effective correlation-length exponent}

\begin{figure}
\putinscaledfigure{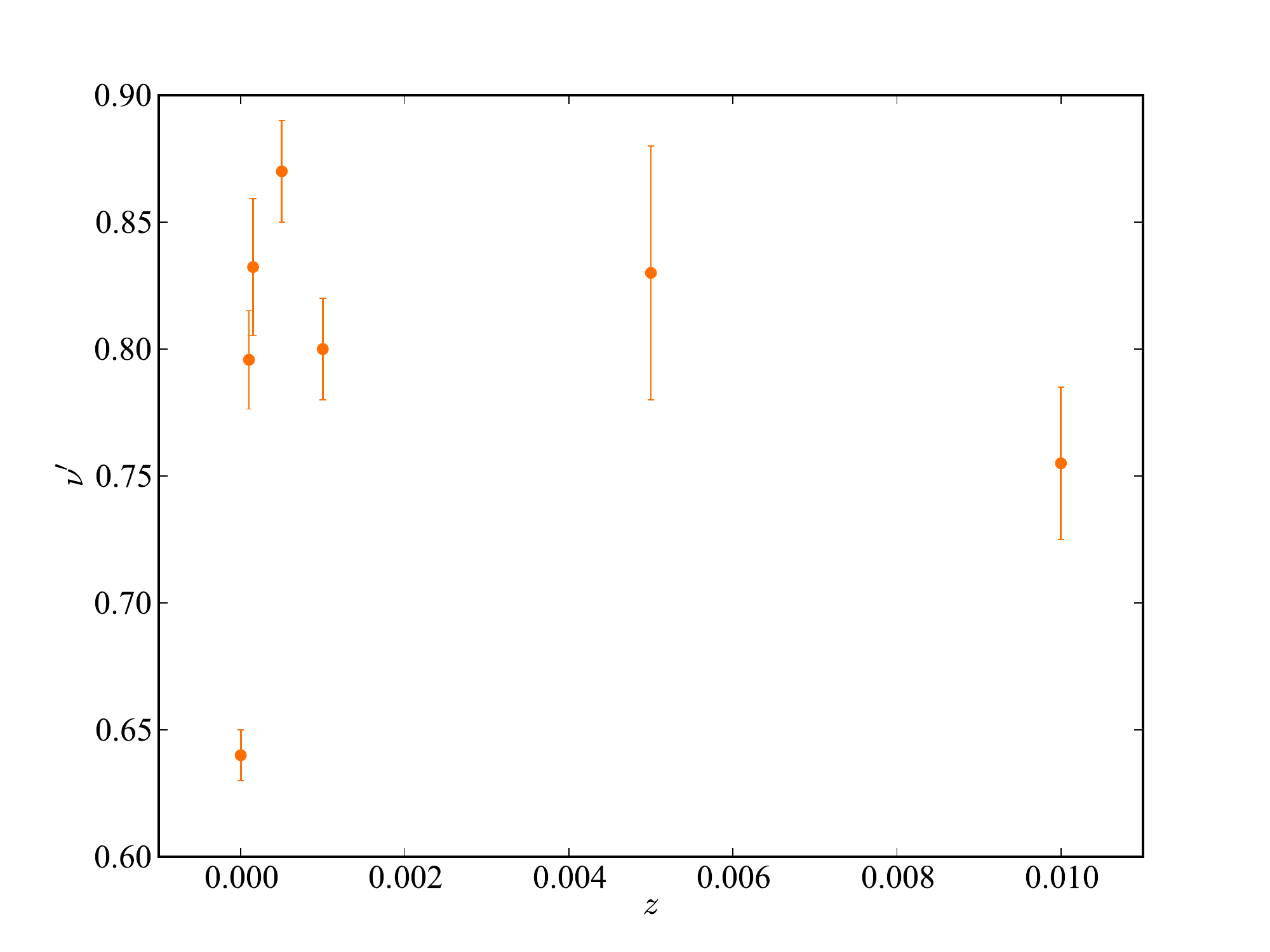}
\caption{Effective correlation-length exponent $\nu'$ as a function of monomer fugacity $z$. For each value of $z$, the correlation-length exponent is found using the temperature-derivative of $\scB$ at the critical temperature, as in \reffig{FigBinderSlopez0}.
\label{FigBinderSlopeOD}}
\end{figure}

The effective value of the correlation-length critical exponent $\nu'$, determined using the slope of $\scB$ at $T\sub{c}(z)$, is shown as a function of $z$ in \reffig{FigBinderSlopeOD}.
While a clear distinction is seen between the exponents for $z=0$ and $z>0$, confirming the different universality classes for these two cases, it is difficult to draw further conclusions about the nature of the transition at $z>0$.

According to the scenario outlined in \refsecand{SecScalingOD}{SecBinderSlope}, one would expect $\nu'$ to take the value\cite{CampostriniHeisenberg} $\nu\sub{Heis.} = 0.7112 \pm 0.0005$ appropriate to the 3D Heisenberg universality class, at least for intermediate values of $z$: For very small $z$, the influence of the confinement fixed point gives corrections (for even quite large $L$), described by \refeq{EqBinderSlopeOD}; for larger $z$ values the true first-order nature of the transition should become apparent, causing complete breakdown of scaling. The large uncertainty in $T\sub{c}(z)$, caused by slow convergence of the crossings of the Binder cumulant, makes refinement of $\nu'$ difficult.

\section{Conclusions}
\label{SecConclusions}

Our central result, illustrated in \reffig{FigTcz}, is the quantitative relationship, derived using the scaling theory and confirmed in large-scale Monte Carlo simulations, between the monopole distribution function at $z=0$ and the phase boundary at $z>0$. The connection between the two follows from their common dependence on the RG eigenvalue $y_z$ of the monomer fugacity at the fixed point describing the confinement transition. There is no reason to expect such a relationship for a conventional order--disorder transition, and so this result provides clear and direct evidence for the claim of a non-Landau transition characterized by confinement.

As illustrated in \reffig{FigBinderSlopeOD}, we also directly observe a difference between the correlation-length critical exponents at the confinement ($z=0$) and order--disorder ($z>0$) transitions. This is a direct demonstration of the distinct critical behavior in the generic case, where Landau theory is expected to apply, and in the constrained limit, where it is violated. Related previous examples include the unconventional critical behavior observed in a topologically constrained Heisenberg model\cite{Motrunich} and recent studies of the square-lattice dimer model at finite hole density.\cite{Papanikolaou2}

Consistently with previous work on the CDM, we see no sign of first-order behavior for $z=0$ and $v_4 \ge 0$.\cite{Alet,Misguich,Charrier2,Charrier,Chen,Papanikolaou} As discussed in \refsec{SecScalingOD}, we expect a very weak first-order transition for small $z>0$, with Heisenberg-like critical behavior on moderate length scales. Our results at $z>0$ and $v_4 = +1$ are indeed broadly consistent with a continuous transition in the Heisenberg universality class, including (1) the absence of a double-peak structure in the energy histogram (\reffig{FigHistograms}), (2) well-defined crossing points of the Binder cumulant (\reffig{FigBinderOD1}), and (3) good fits to the finite-size behavior of the Binder crossings, using the exponents of the Heisenberg universality class (\reffig{FigBinderCrossingsOD}). We also observe reasonable finite-size scaling of the temperature-derivative of the Binder cumulant, but we are unable to determine the exponent $\nu'$ accurately enough to identify the universality class (\reffig{FigBinderSlopeOD}).

The scaling arguments of \refsec{SecScalingOD} indicate that the putative first-order transition at $z>0$ may never be visible with realistic system sizes. Given the substantial corrections to scaling exhibited by this transition, however, such arguments should not be considered definitive. Indeed, at $v_4 = 0$, where corrections due to the tricritical point at $v_4 = v_{4\text{c}}$ are significant, clear first-order behavior is seen at $z=10^{-2}$. More work, including studies at intermediate $v_4$, are needed to clarify this picture.

\subsection*{Universality and deconfined criticality}

The zero-temperature ordering transition in certain $2$D quantum antiferromagnets, such as the [$\mathrm{SU}(2)$] JQ model,\cite{SandvikJQ} is claimed\cite{Senthil} to be described by the same critical theory as given in \refeq{EqCriticalAction}, and should therefore belong in the same universality class. Universality is one of the most remarkable features of critical behavior, and a clear demonstration in the context of non-Landau criticality would be a striking result. Since the confinement transition in the dimer model necessarily occurs at zero monomer density, it would also provide direct evidence for deconfinement at the quantum critical point.

In the JQ model, the formation of a valence-bond solid (VBS) is a consequence of the condensation of monopoles.\cite{Senthil} The exponent $\eta\sub{VBS}$ describing the appearance of VBS order is therefore related to the monopole scaling dimension $y_z$ by $\eta\sub{VBS} = d + 2 - 2y_z$. Values for $y_z$ and for the critical exponent $\nu$, displayed in \reftab{TabExponents}, show reasonable agreement, albeit with large error bars.
\begin{table}
\caption{\label{TabExponents}Comparison of critical exponents for the cubic dimer model and the JQ model\cite{SandvikJQ} with $\mathrm{SU}(2)$ symmetry, on square and honeycomb lattices. For the CDM, two values of $\nu$ are reported in each case, calculated using the Binder cumulant $\scB$ and the stiffness $K$. In this work, two values of $y_z$ are found, using the test-monopole distribution $G\sub{m}$ at $z=0$ and the phase boundary $T\sub{c}(z>0)$. For the JQ model, the monopole scaling dimension $y_z$ is found using the anomalous dimension of the VBS operator\cite{Senthil} $\eta\sub{VBS} = d + 2 - 2y_z$ (denoted $\eta_d$ in \refcite{Lou} and $\eta_\Psi$ in \refcite{Harada}). The two sets of values for the JQ model on the square lattice correspond to two choices for the Q term.\cite{Lou} In \refcite{Harada} good finite-size scaling is observed for both square and honeycomb lattices (for system sizes $L \le 96$), but no uncertainties are quoted.}
\begin{ruledtabular}
\begin{tabular}{lcccc}
&Model&$\nu$&$\eta\sub{VBS}$&$y_z$
\\\hline
\multirow{2}{*}{this work}&\multirow{2}{*}{CDM ($v_4 = +1$)}&$0.64(1)$ ($\scB$)&&$2.421(8)$ ($G\sub{m}$)\\
&&$0.61(1)$ ($K$)&&$2.28(13)$ ($T\sub{c}$)\\
\multirow{2}{*}{\refcite{Charrier2}}&\multirow{2}{*}{CDM ($v_4 = +1$)}&$0.60(4)$ ($\scB$)&&\\
&&$0.61(4)$ ($K$)&&\\\hline
\multirow{2}{*}{\refcite{Lou}}&\multirow{2}{*}{JQ (square)}&$0.67(1)$&0.20(2)&$2.40(1)$\\
&&$0.69(2)$&$0.20(2)$&$2.40(1)$\\
\refcite{Pujari}&JQ (honeycomb)&$0.54(5)$&$0.28(8)$&$2.36(4)$\\
\refcite{Harada}&JQ&0.59&0.35&2.33
\end{tabular}
\end{ruledtabular}
\end{table}

While quantitative comparison of exponents is always challenging, it is made even more so in this case by the presence of large corrections to scaling. Significant corrections are observed in all cases,\cite{Sandvik} providing a further, qualitative indication of universality. This indicates the presence of a weakly irrelevant scaling variable at the fixed point, an observation consistent with a recent functional-RG study\cite{Bartosch} of the critical action \refeq{EqCriticalAction}, which found a small correction-to-scaling exponent, $\omega \simeq 0.3$.

Although scaling behavior is visible over a range of length scales,\cite{KunChen} the possibility remains that the transition in the JQ model is first order.\cite{Kuklov2008} (This should not be confused with the weak first-order transition expected in the CDM at $z>0$. Berry phases suppress singly-charged monopoles in the JQ model,\cite{Senthil} so the transition is effectively at zero fugacity, $z=0$.) Indeed, a drift in the effective exponent values with increasing system sizes was observed by Harada et al.,\cite{Harada} consistent with a weak first-order transition. Universality between different models and different lattices, as suggested by \reftab{TabExponents}, would nonetheless imply that the RG flow is through the neighborhood of a critical fixed point. In this case, one might expect that, as in the CDM,\cite{Papanikolaou,Charrier2} additional perturbations could drive the JQ transition truly continuous, or at least extend the length scale over which scaling can be observed.

\acknowledgments

The simulations were performed on resources provided by the Swedish National Infrastructure for Computing (SNIC) at the National Supercomputing Centre (NSC) and High Performance Computing Center North (HPC2N).

\appendix*

\section{Monte Carlo algorithm}

In this Appendix, we give a brief description of the numerical methods used for the calculation.

The Monte Carlo routine is based on the directed loop algorithm,\cite{SandvikMoessner} adapted to allow for violations of dimer constraints. The system consists of a three dimensional cubic lattice with $L$ sites, or nodes, in each direction and periodic boundary conditions. The dimers can occupy the links of the lattice subject to the dimer constraints described in \refsec{SecModel}. The Monte Carlo algorithm samples the equilibrium thermodynamic distribution of dimer configurations $\propto\exp -\frac{E}{T}$ where the energy $E$ of a configuration is given by \refeq{EqConfigEnergy}.

Two kinds of updates are used for sampling the configurations; the first has a high acceptance rate but does not alter the number of monomers, while the second can generate and annihilate monomers.

\begin{figure}
\putinscaledfigure{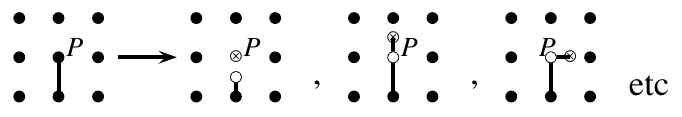}
\caption{First stage of a Monte Carlo update. After a node $P$ is selected, a sequence of dimer rearrangements is performed, starting with creation of a pair of monomers of opposite charges---one on $P$ and one on a link connected to $P$. Half a dimer is added or removed depending on the occupancy of the link.}
\label{fig:start_0}
\end{figure}

The first kind of update begins by randomly picking a monomer-free node $P$. The node is accepted with a probability (specified below) that depends only on the local configuration around $P$. If accepted, the system transitions into a configuration in which there are two monomers---one on the node $P$ and one on a link connected to $P$. This is accompanied by addition or removal of half a dimer as shown in the example in \reffig{fig:start_0}.
The link monomer has a ``direction'' attribute, which serves to identify one of the two nodes connected to the link as the node ``ahead'', and which initially points away from $P$.

\begin{figure}
\putinscaledfigure{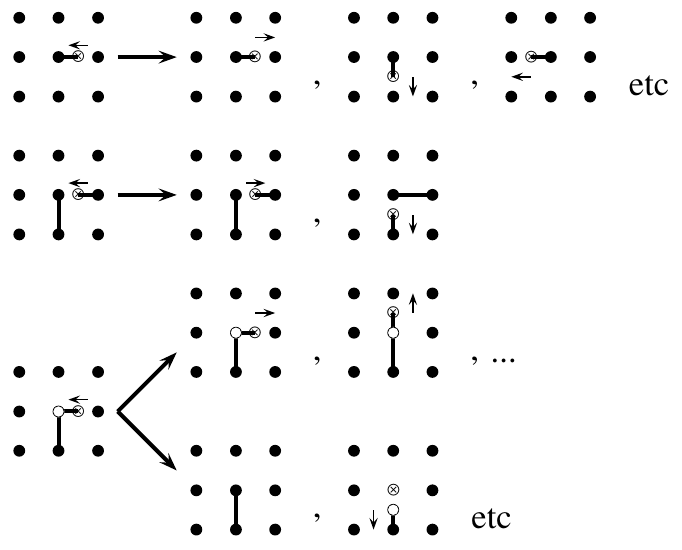}
\caption{Examples of propagation of link monomer. Monomer propagation in the first kind of update proceeds without changing the monomer count on the nodes. If the node ahead has a monomer on it (bottom), the update sequence can terminate with the link-monomer annihilating the one on the node.}
\label{fig:propagation_0}
\end{figure}

The link monomer can hop to one of the six links connected to the node $P_1$ ahead of it, erasing or creating dimers in the process as shown in \reffig{fig:propagation_0}.
Any such hopping that does not result in a change in the number of monomers on $P_1$ is allowed. If the link monomer hops back to the same edge, the direction of propagation is flipped. If $P_1$ has a monomer, the link monomer can annihilate with the one on the node, thereby terminating the update process.

The probability of a transition from configuration $k$ to $q$ is given by 
\beq{EqProbktoq}
p(k\to q) = \frac{\bar{w}_q-\delta_{kq} \min(\bar{w})}{\sum \bar{w} - \min(\bar{w})}\punc{,}
\eeq
where $\bar{w}_i=\exp\left(-\frac{\bar{E}_i}{T}\right)$ with $\bar{E}_i$ the energy of the configuration without considering the two monomers created by the update process. The energy of the half dimer is taken to be half that of a full dimer at the same edge.

The second kind of update is similar to the first, but can start on any node, including ones with monomers. Such an update can terminate with the absorption of the link monomer into a node, creating or annihilating a node monomer. The propagation of the link monomer proceeds similarly to the first update, with probability of transition from $k$ to $q$ of $p(k\to q) = \frac{w_q-\delta_{kq} \min(w)}{\sum w - \min(w)}$ where $w_i=\exp\left(-\frac{E_i}{T}\right)$. Since the number of monomers can change after the update, the energy $E$ is calculated with the energy of every node monomer added by the process.

The relative frequency of the two kinds of update was chosen to minimize the convergence time but otherwise did not affect the outcomes. Thermalization of the system from a monomer-free, fully ordered configuration was achieved by performing the updates until approximately $6000\times L^3$ forward-propagation steps of the link monomer had been achieved. Configuration samples were subsequently taken once every $\sim L^3$ forward propagation steps. Averages and error estimates of thermal expectation values and their combinations (e.g., Binder cumulants) were obtained by the blocking method. 

The crossing point of Binder cumulants $\mathcal{B}(L_1,T)$ and $\mathcal{B}(L_2,T)$ for systems of size $L_1$ and $L_2$ was obtained by first finding an approximate crossing point $\tilde{T}_\times$ from a piecewise linear interpolation of the data. A better estimate $T_\times$ was obtained by fitting the data in the temperature interval $\tilde{T}_\times\pm 0.00375$ to quadratics,
\beq{EqQuadratics}
\begin{aligned}
\mathcal{B}(L_1,T) &= \scB_0 + M_1\left(T-T_\times\right) + A_1\left(T-T_\times\right)^2\\
\mathcal{B}(L_2,T) &= \scB_0 + M_2\left(T-T_\times\right) + A_2\left(T-T_\times\right)^2
\end{aligned}
\eeq
(with fit parameters $\scB_0$, $A_{1,2}$, $M_{1,2}$, and $T_\times$).

The $T$-derivative of the binder cumulant was found using a fluctuation--response relation,
\newcommand{\Mean}[1]{\langle #1 \rangle}
\newcommand{\Mii}{{\lvert\mv\rvert^2}}
\newcommand{\Miv}{{\lvert\mv\rvert^4}}
\begin{multline}
\parderat{\scB}{T}{z} =-\frac{1}{3T^2} \frac{\Mean\Miv}{\Mean\Mii^2} \left[ \frac{\Mean{(E\sub{dimer}-\Mean{E\sub{dimer}})\Miv}}{\Mean\Miv}\right.\\
\left.{}-2\frac{\Mean{(E\sub{dimer}-\Mean{E\sub{dimer}})\Mii}}{\Mean\Mii}  \right]\punc{,}
\label{BinderSlopeEq}
\end{multline}
where $E\sub{dimer}$ is the energy excluding the contribution from monomers.

The Monte Carlo algorithm and its implementation were tested using comparisons with previously published results and, for small system sizes, simple alternative methods. For $L=2$, results were compared with the exact averages obtained from the enumeration of all possible microscopic states. For system size $L=4$ and parameters $v_4=0$ and $z=0.05$, the energy and susceptibility were compared with an elementary Metropolis algorithm. In addition, the self consistency of the slope of the Binder cumulants estimated using \refeq{BinderSlopeEq} and from curve fitting through Binder cumulants supports the validity of the sampling algorithm.


\begin{thebibliography}{99}
\bibitem{FootnoteDualLanguage} The dual language is used in \refcite{Chen}, so our electric charges are there referred to as magnetic monopoles.
\bibitem{Footnotedivm} A term $(\del\cdot\mv)^2$ is also consistent with symmetry, but is irrelevant at the Heisenberg fixed point.\cite{Carmona}
\bibitem{FootnoteIsotropic} This result is asymptotically correct for $R$ large compared to the lattice scale. The form of \refeq{EqCriticalAction} implies that $G\sub{m}$ is then isotropic and sublattice independent.
\bibitem{FootnoteFlux} For $z=0$, the lattice Gauss law implies that $\phi_\mu(R)$ is independent of $R$ and so the sum over $R$ is redundant.\cite{Alet}
\bibitem{FootnoteBoundTc} Reduced size (with periodic boundaries) tends to increase the ordering temperature, and hence the Binder crossing point, because it restricts fluctuations to those commensurate with the system period.\cite{Cardy} The stiffness is sensitive to only those fluctuations that change the total magnetic flux through the system. Threading a unit of flux is easier for a smaller system, and so one needs lower temperature to suppress such fluctuations for smaller $L$.
\end{thebibliography}
\end{document}